\documentclass[a4paper,11pt]{article}
\usepackage{jheppub} 
\usepackage{lineno}
\usepackage{graphicx}
\usepackage{dcolumn}
\usepackage{bm}
\usepackage{appendix}
\usepackage{slashed}

\title{Right-Handed Neutrino Production by an Axion-like Inflaton: Implications for Leptogenesis}

\begin{document}

\author[a]{Weiyi Deng}
\author[a,b]{Chengcheng Han}
\author[a]{Wuzhou Yin}
\author[a]{Tong Ju}

\affiliation[a]{School of Physics, Sun Yat-Sen University, Guangzhou 510275, P. R. China}
\affiliation[b]{Asia Pacific Center for Theoretical Physics, Pohang 37673, Korea}

\emailAdd{dengwy23@mail2.sysu.edu.cn}
\emailAdd{hanchch@mail.sysu.edu.cn}
\emailAdd{yinwzh3@mail2.sysu.edu.cn}
\emailAdd{jutong@mail2.sysu.edu.cn}

\abstract{
We study heavy right-handed neutrino production induced by a derivative coupling to an axion-like inflaton and its implications for non-thermal leptogenesis. We develop a unified treatment of Majorana-fermion production during inflation and preheating. During inflation, the rolling inflaton background generates a helicity-asymmetric spectrum that can be obtained analytically in the slow-roll regime, while during preheating the oscillating background drives repeated non-adiabatic production events with a helicity-dependent effective momentum and a natural momentum cutoff. We compute the resulting abundance in both the delayed-decay and prompt-decay limits, including the effects of Pauli blocking and successive production stages. We further clarify the relation between the fermion basis used to identify non-adiabatic production and the Hamiltonian-diagonal basis used to define instantaneous occupation numbers. Applying these results to non-thermal leptogenesis, we identify parameter regions consistent with the observed baryon asymmetry.
}

\maketitle
\flushbottom

\section{Introduction}
\label{sec:intro}

The observed baryon asymmetry of the Universe~\cite{Planck:2018vyg} and the nonzero masses of active neutrinos~\cite{Super-Kamiokande:1998kpq,SNO:2002tuh,ParticleDataGroup:2024cfk} provide two important hints of physics beyond the Standard Model.
A minimal and well-motivated framework that can address both problems is the type-I seesaw mechanism with right-handed neutrinos~\cite{Minkowski:1977sc,Yanagida:1979as,Yanagida:1980xy,Gell-Mann:1979vob,Mohapatra:1979ia,Mohapatra:1980yp,Schechter:1980gr,Schechter:1981cv}.
In this framework, the smallness of active neutrino masses is explained by the large Majorana masses of right-handed neutrinos, while the CP-violating decays of these heavy states can generate a primordial lepton asymmetry~\cite{Sakharov:1967dj,Fukugita:1986hr}, which is subsequently converted into the baryon asymmetry through electroweak sphaleron processes~\cite{Kuzmin:1985mm,Harvey:1990qw}.

In the canonical seesaw scenario, Yukawa couplings of order unity, as often expected in grand unified theories, typically point to a seesaw scale around $10^{13}$--$10^{14}\,{\rm GeV}$.
In conventional thermal leptogenesis~\cite{Luty:1992un,Plumacher:1996kc,Barbieri:1999ma,Giudice:2003jh,Buchmuller:2004nz,Buchmuller:2005eh,Garbrecht:2018mrp}, producing such heavy states from the thermal bath requires a reheating temperature comparable to the seesaw scale.
This requirement can be restrictive, especially when the reheating temperature is lower or when thermal production is otherwise inefficient.

This motivates production mechanisms that do not rely on a thermal bath. In particular, the seesaw scale can be close to the Hubble scale during high-scale inflation.
It is therefore natural to consider the possibility that heavy right-handed neutrinos are produced non-thermally from the inflaton sector.
Such production may occur perturbatively through inflaton decays~\cite{Lazarides:1990huy,Asaka:1999jb,Asaka:1999yd,Asaka:2002zu,Senoguz:2003hc,Hahn-Woernle:2008tsk,Buchmuller:2013dja,Barman:2021tgt,Barman:2024ujh,Han:2024qbw,You:2024hit}, or non-perturbatively through the time-dependent inflaton background and the subsequent preheating dynamics~\cite{Baacke:1998di,Felder:1998vq,Greene:1998nh,Giudice:1999fb,Peloso:2000hy,Adshead:2015jza,Adshead:2015kza,Adshead:2018oaa,Kulkarni:2024kzc}.
If right-handed neutrinos couple directly to the inflaton, they can already be generated during inflation~\cite{Adshead:2015jza,Adshead:2015kza,Adshead:2018oaa} and may be further produced during preheating~\cite{Felder:1998vq,Greene:1998nh,Giudice:1999fb,Peloso:2000hy,Adshead:2015jza,Adshead:2015kza}.
The resulting non-thermal population can subsequently decay through the standard leptogenesis mechanism and source the observed baryon asymmetry, even when thermal production is inefficient or absent.
This establishes a direct connection among early-Universe particle production, neutrino mass generation, and baryogenesis.

In this work, we consider a concrete realization in which an axion-like inflaton is derivatively coupled to a fermion current~\cite{Adshead:2015kza,Adshead:2015jza,Kaloper:2008qs,Adshead:2018oaa,Kulkarni:2024kzc}.
The rolling inflaton background then acts as an effective time-dependent axial background and can lead to helicity-dependent fermion production.
This provides a natural setting for connecting fermion production during inflation and preheating with non-thermal leptogenesis, especially when the produced fermions are identified with heavy right-handed neutrinos.
Compared with bosonic preheating~\cite{Kofman:1994rk,Shtanov:1994ce,Kitajima:2017peg}, fermionic production is constrained by Pauli blocking, but it can still be efficient when the evolution becomes non-adiabatic.
A distinctive feature of the derivative coupling is the helicity asymmetry imprinted on the produced spectrum, which can play an important role in the subsequent generation of a lepton asymmetry.
We therefore follow the production history from inflation to preheating and finally to the decay of the produced fermions.
In applying this mechanism to leptogenesis, we distinguish between delayed-decay and prompt-decay regimes, according to whether the produced fermions survive over many inflaton oscillations or decay shortly after each production event.

Our analysis goes beyond previous studies in several respects. First, we treat the production of Majorana fermions during inflation and preheating within a single framework and follow the full production history of the non-thermal population. In particular, the helicity-asymmetric spectrum generated during inflation is not regarded as an isolated result, but is used as the initial condition for the subsequent preheating evolution whenever the produced fermions are sufficiently long-lived. Second, we emphasize that the final abundance depends qualitatively on the lifetime of the produced fermions. In the delayed-decay regime, the occupation number accumulates over many inflaton oscillations and is eventually limited by Pauli blocking, whereas in the prompt-decay regime each non-adiabatic burst acts as an independent source and the relevant abundance is obtained by summing the stage-by-stage yields. This distinction has a direct impact on the resulting baryon asymmetry. Third, we clarify the relation between the fermion bases commonly used in this problem. While the original $Y$ basis makes the origin of non-adiabaticity transparent through the helicity-dependent effective momentum, the instantaneous particle number is defined in the Hamiltonian-diagonal $\psi$ basis. We show that this distinction is essential, especially in the small-mass regime, where a naive occupation number defined in the $Y$ basis can lead to basis-induced, unphysical behavior. These points allow us to connect axion-induced fermion production with non-thermal leptogenesis in a controlled and physically transparent manner.

The rest of this paper is organized as follows.
In section~\ref{sec.2}, we introduce the model and set up the formalism for fermion production.
In section~\ref{sec:fermion_production}, we analyze fermion production during inflation and preheating, including both delayed-decay and prompt-decay regimes.
In section~\ref{sec.4}, we apply the resulting non-thermal abundance to leptogenesis from right-handed neutrino decays.
We conclude in section~\ref{sec:conclusion}.

\section{Framework: fermion bases and instantaneous occupation number}
\label{sec.2}

In this section, we present the general framework for Majorana-fermion production in a time-dependent background. Our discussion closely follows the standard Bogolyubov treatment of fermion production in such backgrounds~\cite{Adshead:2015kza,Adshead:2018oaa,Kulkarni:2024kzc}.

Our goal is to establish a formulation that makes both the origin of non-adiabatic particle production and the definition of particle number manifest. To this end, we first introduce the model and two useful fermion bases that will be employed throughout the paper. We then define the instantaneous occupation number in the Hamiltonian-diagonal $\psi$ basis. This definition is not a basis-independent observable in a time-dependent background, but it provides the physically appropriate particle number used in the abundance calculation.
The two field bases describe the same physical theory. We use the $Y$ basis to make the origin of non-adiabaticity transparent, while the instantaneous particle number is defined in the Hamiltonian-diagonal $\psi$ basis. 

Throughout this work, we distinguish between field-basis equivalence and the operational definition of instantaneous particle number. The $Y$ and $\psi$ bases are related by a time-dependent chiral rotation and therefore describe the same physical theory. However, in a time-dependent background the instantaneous occupation number depends on the choice of positive- and negative-frequency basis. We use the $Y$ basis to make the helicity-dependent non-adiabaticity transparent, while the fermion abundance entering the leptogenesis calculation is defined by diagonalizing the instantaneous Hamiltonian in the $\psi$ basis.

\subsection{Model and two useful fermion bases}\label{sec.21}

We work in the Weyl representation of the $\gamma$ matrices,
\begin{equation}\label{eq:gamma_weyl}
\gamma^0=
\begin{pmatrix}
0 & 1 \\
1 & 0
\end{pmatrix},
\qquad
\gamma^i=
\begin{pmatrix}
0 & \sigma_i \\
-\sigma_i & 0
\end{pmatrix},
\qquad
\gamma^5=i\gamma^0\gamma^1\gamma^2\gamma^3=
\begin{pmatrix}
-1 & 0 \\
0 & 1
\end{pmatrix}.
\end{equation}
We consider a Majorana fermion $X$ coupled to an axion-like inflaton $\phi$ through a dimension-five derivative operator in a spatially flat FLRW background. The Lagrangian is
\begin{equation}\label{eq:Lagrangian}
\mathcal{L}
=
a^4
\left\{
\bar{X}
\left[
i\left(\frac{\gamma^\mu}{a}\partial_\mu+\frac{3}{2}H\gamma^0\right)
-m
+\frac{1}{f}\frac{\gamma^\mu}{a}\gamma^5\partial_\mu\phi
\right]
X
+\frac{1}{2}(\partial\phi)^2
-
V(\phi)
\right\},
\end{equation}
where $a(t)$ is the scale factor, $H=\dot a/a=a'/a^2$ is the Hubble parameter, and $f$ characterizes the axion--fermion coupling. We work in conformal time $\tau$, with $\partial_0=\partial_\tau$, and use a prime and a dot to denote derivatives with respect to conformal time and cosmic time, respectively. Throughout this work, we neglect metric perturbations and treat the inflaton as a spatially homogeneous classical background, $\phi(\mathbf{x},\tau)\simeq \phi(\tau)$.
It is convenient to introduce the dimensionless field
\begin{equation}
\theta \equiv \frac{\phi}{f},
\end{equation}
and to rescale the fermion field as
\begin{equation}
Y \equiv a^{3/2} X.
\end{equation}
In terms of $Y$, the fermionic part of the Lagrangian becomes
\begin{equation}\label{eq:Y_Lagrangian}
\mathcal{L}\supset
\bar{Y}
\left(
i\gamma^\mu\partial_\mu
-ma
+\gamma^\mu\gamma^5\partial_\mu\theta
\right)
Y,
\end{equation}
and the corresponding equation of motion is
\begin{equation}\label{eq:Y_eom}
\left(
i\gamma^\mu\partial_\mu
-ma
+\gamma^0\gamma^5\theta'
\right)Y=0.
\end{equation}
We begin in the $Y$ basis, where the effect of the time-dependent inflaton background enters explicitly through the derivative coupling. Expanding $Y$ in plane waves, similarly to Eq.~\eqref{eq:psi_FT}, we write
\begin{equation}\label{eq:Y_fourier}
Y(\mathbf{x})
=
\int \frac{\text{d}^3 k}{(2\pi)^{3/2}} e^{i \mathbf{k} \cdot \mathbf{x}}
\sum_{r=\pm}
\left[
\tilde{U}_r(\mathbf{k},\tau) a_r(\mathbf{k})
+
\tilde{V}_r(-\mathbf{k},\tau) a_r^\dagger(-\mathbf{k})
\right],
\end{equation}
where
\begin{equation}\label{eq:UVtilde_mode}
\tilde{U}_r(\mathbf{k}, \tau)
=
\frac{1}{\sqrt{2}}
\begin{pmatrix}
\chi_r(\hat{\mathbf{k}})\, \tilde{u}_r(k, \tau) \\
r\,\chi_r(\hat{\mathbf{k}})\, \tilde{v}_r(k, \tau)
\end{pmatrix},
\qquad
\tilde{V}_r(\mathbf{k}, \tau)
=
\frac{1}{\sqrt{2}}
\begin{pmatrix}
\chi_r(\hat{\mathbf{k}})\, \tilde{w}_r(k, \tau) \\
r\,\chi_r(\hat{\mathbf{k}})\, \tilde{y}_r(k,\tau)
\end{pmatrix}.
\end{equation}
The two-component helicity eigenspinors are defined by
\begin{equation}\label{eq:chi_def}
\chi_r(\hat{\mathbf{k}})
=
\frac{1+r\,\boldsymbol{\sigma}\cdot\hat{\mathbf{k}}}
{\sqrt{2(1+\hat{k}_3)}}
\,\bar{\chi}_r,
\qquad
\hat{\mathbf{k}}=\frac{\mathbf{k}}{|\mathbf{k}|},
\qquad
\bar{\chi}_+ =
\begin{pmatrix}
1\\0
\end{pmatrix},
\qquad
\bar{\chi}_- =
\begin{pmatrix}
0\\1
\end{pmatrix},
\end{equation}
and satisfy
\begin{equation}\label{eq:chi_prop}
\boldsymbol{\sigma}\cdot\hat{\mathbf{k}}\,\chi_r(\hat{\mathbf{k}})
=
r\,\chi_r(\hat{\mathbf{k}}),
\qquad
\chi_r^\dagger(\hat{\mathbf{k}})\chi_s(\hat{\mathbf{k}})
=
\delta_{rs}.
\end{equation}
Using
\begin{equation}
i\sigma_2\chi_r^*(\hat{\mathbf{k}})=-r\,\chi_{-r}(\hat{\mathbf{k}}),
\end{equation}
the charge-conjugation relation implies
\begin{equation}
\tilde w_r=\tilde v_{-r}^*,
\qquad
\tilde y_r=\tilde u_{-r}^*.
\label{ccrelation}
\end{equation}
Substituting eqs.~\eqref{eq:gamma_weyl} and \eqref{eq:Y_fourier} into the equation of motion \eqref{eq:Y_eom}, we obtain
\begin{align}
i{\tilde{u}}_r'&=-(kr - a\dot{\theta})\tilde{u}_r + m a r\tilde{v}_r,
\label{eq:u_eq_Y} \\
i{\tilde{v}}_r'&=(kr - a\dot{\theta})\tilde{v}_r+  m a r\tilde{u}_r,
\label{eq:v_eq_Y}
\end{align}
where a prime denotes a derivative with respect to conformal time.
These equations show that the inflaton background induces a helicity-dependent shift in the effective momentum, $k \to k - r a\dot{\theta}$. As a result, the evolution can become non-adiabatic when this effective momentum approaches zero. This feature will play a central role in the analysis of fermion production in the subsequent sections.

In the $Y$ basis, the role of the time-dependent inflaton background in driving non-adiabatic fermion production is particularly transparent. The derivative coupling $\gamma^\mu\gamma^5\partial_\mu\theta$ enters directly in the equations of motion as a helicity-dependent shift of the effective momentum, making the origin of particle production manifest. However, as emphasized in Ref.~\cite{Adshead:2018oaa}, in the regime $\dot{\theta}\gg H$ the Hamiltonian in this basis does not provide a convenient starting point for a particle interpretation. The derivative coupling acts as a time-dependent axial background, which mixes positive- and negative-frequency modes in a way that obscures the identification of particle states. As a result, the Hamiltonian is not diagonal in the canonical creation and annihilation operators, and the notion of particles becomes ambiguous.

It is therefore useful to perform the chiral rotation
\begin{equation}\label{eq:psi_Y_trans}
\psi \equiv e^{-i\gamma^5\theta}Y.
\end{equation}
In terms of $\psi$, the fermionic Lagrangian becomes
\begin{equation}\label{eq:psi_Lagragian}
\mathcal{L}\supset
\bar{\psi}
\left(
i\gamma^\mu\partial_\mu
-
ma\,e^{2i\gamma^5\theta}
\right)\psi
=
\bar{\psi}
\left\{
i\gamma^\mu\partial_\mu
-
ma
\left[
\cos\left(\frac{2\phi}{f}\right)
+
i\gamma^5\sin\left(\frac{2\phi}{f}\right)
\right]
\right\}\psi,
\end{equation}
Varying the action with respect to $\bar\psi$, we obtain the equation of motion
\begin{equation}\label{eq:psi_eom}
\left(
i\gamma^\mu\partial_\mu
-
ma\,e^{2i\gamma^5\theta}
\right)\psi=0.
\end{equation}

In this basis, the derivative coupling is traded for a time-dependent complex mass term. This form of the Lagrangian leads to a Hamiltonian that can be diagonalized at each instant of time, allowing for a well-defined notion of instantaneous energy eigenstates and a consistent construction of Bogolyubov coefficients. 
The two bases are related by a field redefinition and therefore describe the same physical system. However, the instantaneous particle number is an operational quantity whose definition depends on the choice of positive- and negative-frequency basis. In this work, we use the $Y$ basis only to diagnose the origin of non-adiabaticity, while the fermion abundance is defined through the Bogolyubov coefficients obtained by diagonalizing the Hamiltonian in the $\psi$ basis.

To solve Eq.~\eqref{eq:psi_eom}, and to make contact with the particle interpretation introduced above, we expand the Majorana field in Fourier modes,
\begin{equation}\label{eq:psi_FT}
\psi(\mathbf{x})
=
\int \frac{\mathrm{d}^3k}{(2\pi)^{3/2}}
e^{i\mathbf{k}\cdot\mathbf{x}}
\sum_{r=\pm}
\left[
U_r(\mathbf{k},t)a_r(\mathbf{k})
+
V_r(-\mathbf{k},t)a_r^\dagger(-\mathbf{k})
\right],
\end{equation}
where $a_r(\mathbf{k})$ and $a_r^\dagger(\mathbf{k})$ are the annihilation and creation operators for a Majorana fermion of helicity $r$. 
This expansion corresponds to a decomposition into positive- and negative-frequency modes, which will later be related to instantaneous quasiparticle excitations through the Bogolyubov transformation.
The mode functions are parametrized as
\begin{equation}\label{eq:psi_mode_f}
U_r(\mathbf{k},t)
=
\frac{1}{\sqrt{2}}
\begin{pmatrix}
\chi_r(\hat{\mathbf{k}})\,u_r(k,t) \\
r\,\chi_r(\hat{\mathbf{k}})\,v_r(k,t)
\end{pmatrix},
\qquad
V_r(\mathbf{k},t)
=
\frac{1}{\sqrt{2}}
\begin{pmatrix}
\chi_r(\hat{\mathbf{k}})\,w_r(k,t) \\
r\,\chi_r(\hat{\mathbf{k}})\,y_r(k,t)
\end{pmatrix},
\end{equation}
where the helicity eigenspinors $\chi_r(\hat{\mathbf{k}})$ are defined in Eq.~\eqref{eq:chi_def}. This parametrization isolates the helicity structure and reduces the dynamics to the time-dependent mode functions $u_r$, $v_r$, $w_r$, and $y_r$. The Majorana condition relates the positive- and negative-frequency mode functions through charge conjugation,
\begin{equation}
V_r(\mathbf{k}) = C \bar U_r^T(\mathbf{k}),
\qquad
C=i\gamma^0\gamma^2,
\end{equation}
which relates the two sets of mode functions. As a result, not all components are independent: the degrees of freedom can be reduced to $u_r$ and $v_r$, with $w_r$ and $y_r$ fixed by the charge-conjugation condition.

The independent mode functions satisfy the normalization condition
\begin{equation}
|u_r|^2+|v_r|^2=2,
\end{equation}
and evolve according to
\begin{equation}
\begin{aligned}
i\dot{u}_r + r\frac{k}{a}u_r - m r e^{2i\theta}v_r &= 0\, , \\
i\dot{v}_r - r\frac{k}{a}v_r - m r e^{-2i\theta}u_r &= 0\, ,
\end{aligned}
\label{eq:u_v_general}
\end{equation}
These equations explicitly show that the time-dependent background $\theta(t)$ induces a mixing between $u_r$ and $v_r$. This mixing reflects the non-conservation of particle number in a time-dependent background and provides the microscopic origin of fermion production. In particular, the phase factor $e^{\pm 2i\theta}$ encodes the dynamics of the inflaton background, whose evolution controls when and how efficiently this mixing becomes non-adiabatic.

From the Lagrangian \eqref{eq:psi_Lagragian}, the fermion Hamiltonian in the $\psi$ basis is
\begin{equation}
H = \int \mathrm{d}^3x\, \bar{\psi}
\left[
-i\, \gamma^i \partial_i + 
ma\cos2\theta + i\gamma^5\, ma\sin2\theta
\right] \psi,
\label{29}
\end{equation}
which takes the form of a fermion with a time-dependent complex mass.
To define the fermion occupation number, we diagonalize the instantaneous Hamiltonian. In momentum space, the Hamiltonian can be written as
\begin{equation}\label{eq:psi_H}
H
=
\int \mathrm{d}^3k
\sum_{r=\pm}
\begin{pmatrix}
\hat a_r^\dagger(\mathbf{k}) &
\hat a_r(-\mathbf{k})
\end{pmatrix}
\begin{pmatrix}
A_r(k,t) & B_r^*(k,t) \\
B_r(k,t) & -A_r(k,t)
\end{pmatrix}
\begin{pmatrix}
\hat a_r(\mathbf{k}) \\
\hat a_r^\dagger(-\mathbf{k})
\end{pmatrix}.
\end{equation}
Using the helicity-spinor identities
\begin{equation}
i\sigma_2\chi_r^\ast(\hat{\mathbf{k}})=-r\chi_{-r}(\hat{\mathbf{k}}),
\qquad
\chi_r(-\hat{\mathbf{k}})=-r e^{i r \varphi_\mathbf{k}} \chi_{-r}(\hat{\mathbf{k}}),
\label{helicity-spinor identities}
\end{equation}
where $e^{i \varphi_\mathbf{k}} \equiv (k_1 + i k_2) / \sqrt{k_1^2 + k_2^2}$ encodes the azimuthal phase of the momentum, we obtain the following matrix elements of the Hamiltonian: 
\begin{equation}\label{eq:Hamiltonian_A_B}
\begin{aligned}
A_r &\equiv \frac{r}{2} \left[ k (|v_r|^2-|u_r|^2)+ ma (u_rv_r^\ast + u_r^\ast v_r)\cos2\theta+i ma (v_ru_r^\ast - u_rv_r^\ast)\sin2\theta\right], \\
B_r &\equiv \frac{e^{i r \varphi_{\mathbf{k}}}}{2} \left[ ma(v_r^2-u_r^2)\cos2\theta-2ku_rv_r+ima(u_r^2+v_r^2)\sin2\theta\right].
\end{aligned}
\end{equation}
The diagonal component $A_r$ determines the instantaneous energy splitting, while the off-diagonal component $B_r$ encodes the mixing between positive- and negative-frequency modes. In a time-dependent background, this mixing is generically nonzero, and the Hamiltonian is therefore not diagonal in the $(\hat a_r,\hat a_r^\dagger)$ basis. As a consequence, the operators $\hat a_r^\dagger$ and $\hat a_r$ do not create and annihilate instantaneous energy eigenstates.

This non-diagonal structure reflects the non-adiabatic evolution of the system and provides the origin of fermion production. A consistent particle interpretation is recovered by diagonalizing the Hamiltonian via a Bogolyubov transformation, which defines the instantaneous energy eigenstates and the corresponding occupation number. To define a basis associated with instantaneous energy eigenstates, we introduce the Bogolyubov transformation
\begin{equation}\label{eq:BT}
\begin{pmatrix}
\hat A_r(\mathbf{k},t) \\
\hat A_r^\dagger(-\mathbf{k},t)
\end{pmatrix}
=
\begin{pmatrix}
\alpha_r(k,t) & -\beta_r^*(k,t) \\
\beta_r(k,t) & \alpha_r^*(k,t)
\end{pmatrix}
\begin{pmatrix}
\hat a_r(\mathbf{k}) \\
\hat a_r^\dagger(-\mathbf{k})
\end{pmatrix},
\end{equation}
where the Bogolyubov coefficients satisfy the normalization condition
\begin{equation}\label{eq:alpha_beta_normalization}
|\alpha_r|^2+|\beta_r|^2=1.
\end{equation}
This transformation diagonalizes the Hamiltonian,
\begin{equation}\label{eq:diagH}
\begin{pmatrix}
A_r & B_r^* \\
B_r & -A_r
\end{pmatrix}
=
\begin{pmatrix}
\alpha_r^* & \beta_r^* \\
-\beta_r & \alpha_r
\end{pmatrix}
\begin{pmatrix}
\omega_k & 0 \\
0 & -\omega_k
\end{pmatrix}
\begin{pmatrix}
\alpha_r & -\beta_r^* \\
\beta_r & \alpha_r^*
\end{pmatrix},
\end{equation}
with
\begin{equation}\label{eq:Bogolyubov_prop}
|\alpha_r|^2-|\beta_r|^2=\frac{A_r}{\omega_k},
\qquad
2\alpha_r\beta_r=-\frac{B_r}{\omega_k}.
\end{equation}
An explicit solution for the Bogolyubov coefficients is
\begin{equation}
\begin{aligned}
\alpha_{r} &= e^{i r \varphi_{\mathbf{k}} / 2} \left[ \frac{1}{2} \sqrt{1 - \frac{kr}{\omega_k}} \, e^{-i \theta}\, u_r + \frac{r}{2} \sqrt{1 + \frac{kr}{\omega_k}} \, e^{i \theta} \, v_r \right], \\
\beta_{r} &= e^{i r \varphi_{\mathbf{k}} / 2} \left[ \frac{1}{2} \sqrt{1 + \frac{kr}{\omega_k}} \, e^{-i \theta} \, u_r - \frac{r}{2} \sqrt{1 - \frac{kr}{\omega_k}} \, e^{i \theta} \, v_r \right],
\end{aligned}
\label{eq:alpha_beta_general}
\end{equation}
where
\begin{equation}
\omega_k = \sqrt{k^2+m^2a^2}.
\end{equation}
The phase $\varphi_{\mathbf{k}}$ reflects the residual phase convention in the definition of the helicity spinors and hence of the Bogolyubov coefficients.
While $\alpha_r$ and $\beta_r$ depend on this residual phase convention, the occupation number $|\beta_r|^2$ defined in the same instantaneous basis is independent of this convention.

Combining eqs.~\eqref{eq:Y_fourier}, \eqref{eq:UVtilde_mode}, and \eqref{eq:psi_Y_trans}, the chiral rotation from the $Y$ basis to the $\psi$ basis induces the following relation between the mode functions:
\begin{equation}\label{eq:uvtil}
u_r(k,t)=e^{i\theta(t)}\tilde{u}_r(k,t),
\qquad
v_r(k,t)=e^{-i\theta(t)}\tilde{v}_r(k,t).
\end{equation}
Substituting this relation into the general expression for the Bogolyubov coefficients in Eq.~\eqref{eq:alpha_beta_general}, one obtains
\begin{equation}
\begin{aligned}
\alpha_{r} &= e^{i r \varphi_{\mathbf{k}} / 2} \left[ \frac{1}{2} \sqrt{1 - \frac{kr}{\omega_k}}\, \tilde{u}_r + \frac{r}{2} \sqrt{1 + \frac{kr}{\omega_k}}\, \tilde{v}_r \right], \\
\beta_{r} &= e^{i r \varphi_{\mathbf{k}} / 2} \left[ \frac{1}{2} \sqrt{1 + \frac{kr}{\omega_k}}\, \tilde{u}_r - \frac{r}{2} \sqrt{1 - \frac{kr}{\omega_k}}\, \tilde{v}_r \right].
\end{aligned}
\label{eq:alpha_beta_inflation}
\end{equation}
This form will be particularly useful for the inflationary calculation, where the mode functions are most conveniently solved in the $Y$ basis.

As a result, the Hamiltonian takes the diagonal form
\begin{equation}\label{eq:diag_H}
H
=
\int \mathrm{d}^3k
\sum_{r=\pm}
\omega_k(t)
\left[
\hat A_r^\dagger(\mathbf{k},t)\hat A_r(\mathbf{k},t)
-
\hat A_r(-\mathbf{k},t)\hat A_r^\dagger(-\mathbf{k},t)
\right],
\end{equation}
which is expressed in terms of operators that create and annihilate instantaneous energy eigenstates.
The instantaneous fermion occupation number for a mode with momentum $\mathbf{k}$ and helicity $r$ is then defined as
\begin{equation}\label{eq:occ_n_k}
\langle 0|
\hat A_r^\dagger(\mathbf{k},t)\hat A_r(\mathbf{k},t)
|0\rangle
=
|\beta_r(\mathbf{k},t)|^2,
\end{equation}
where $|0\rangle$ denotes the initial vacuum state. This quantity measures the excitation relative to the instantaneous particle basis selected by diagonalizing the $\psi$-basis Hamiltonian. The bound $|\beta_r|^2\le 1$ reflects the Pauli exclusion principle.

Integrating over phase space gives the total particle number in a comoving volume,
\begin{equation}\label{eq:N_k}
N_r(t)
=
\sum_{\mathbf{k}}|\beta_r(\mathbf{k},t)|^2
=
\frac{V_{\rm com}}{(2\pi)^3}
\int \mathrm{d}^3k\,
|\beta_r(\mathbf{k},t)|^2,
\end{equation}
and the corresponding physical number density is
\begin{equation}\label{eq:n_k}
n_r(t)
\equiv
\frac{N_r(t)}{V_{\rm phy}(t)}
=
\frac{1}{a^3(t)}
\int\frac{\mathrm{d}^3k}{(2\pi)^3}
|\beta_r(\mathbf{k},t)|^2.
\end{equation}
This expression automatically incorporates the dilution due to cosmic expansion.

Before proceeding, we comment briefly on the distinction between Dirac and Majorana fermions. While their operator structures and mode expansions differ, the independent mode functions satisfy identical equations of motion, and the resulting occupation number $|\beta_r|^2$ is the same after the Bogolyubov transformation~\cite{Adshead:2015kza}. Therefore, the production mechanism discussed here is the same for the two cases at the level of the independent mode functions and single-particle spectrum, with differences arising only in the interpretation of the produced states in specific models.

In this section, we have established a field-redefinition-consistent framework for fermion production in a time-dependent background. The two fermion bases describe the same theory, but they serve different purposes: the $Y$ basis makes the non-adiabaticity transparent, while the $\psi$ basis provides the Hamiltonian-diagonal definition of the instantaneous occupation number used in the abundance calculation. The resulting Bogolyubov coefficient $|\beta_r(k,t)|^2$ encodes the occupation number of fermionic modes and serves as the central quantity characterizing particle production. Its time evolution is governed by the mixing between positive- and negative-frequency modes, as reflected in the off-diagonal structure of the Hamiltonian. A more detailed discussion of why the Hamiltonian-diagonal $\psi$ basis provides the appropriate definition of the instantaneous occupation number, and why the corresponding definition in the $Y$ basis can become ill behaved in the small-mass regime, is given in Appendix~\ref{app1}.

In the following sections, we will analyze how this mixing arises dynamically in different regimes and identify the physical conditions under which fermion production becomes efficient. In particular, we will relate the non-adiabatic evolution to an effective momentum scale that provides an intuitive understanding of the production mechanism.

\section{Production history during inflation and preheating}
\label{sec:fermion_production}

Having established the framework for defining fermion occupation numbers, we now turn to the dynamical production of Majorana fermions in the time-dependent background of an axion-like inflaton. The central quantity is the occupation number $|\beta_r(k,t)|^2$ defined in the Hamiltonian-diagonal $\psi$ basis, whose evolution encodes the non-adiabatic mixing between positive- and negative-frequency modes. Our goal is to determine the helicity-dependent occupation number, the characteristic momentum scale of the produced fermions, and the resulting number density in the relevant cosmological regimes.

We organize the analysis according to the physical production history. During inflation, the slow-roll evolution of the inflaton induces a quasi-adiabatic but helicity-dependent deformation of the fermion spectrum, which admits analytic control and makes the origin of helicity asymmetry particularly transparent. In contrast, during preheating the oscillating inflaton background drives repeated non-adiabatic evolution, leading to burst-like particle production events.

A qualitatively new aspect arises in the preheating regime: the role of the fermion lifetime. If the produced fermions remain in the system, Pauli blocking progressively suppresses further production and leads to a saturated distribution. If instead the fermions decay rapidly, the relevant quantity is the cumulative production from successive non-adiabatic events, with each burst contributing independently.

The two regimes correspond to distinct physical limits of fermion production. In the delayed-decay preheating regime, the produced fermions accumulate in phase space and Pauli blocking drives the system toward a saturated non-thermal distribution. In contrast, in the prompt-decay regime the produced fermions do not persist, and particle production proceeds as a sequence of independent non-adiabatic events, with the total abundance determined by their cumulative contribution. The latter limit is particularly relevant for baryogenesis scenarios, where each production event acts as a localized source. The production dynamics derived here will be used directly in section~4, where they are mapped onto the right-handed neutrino framework.

\subsection{Fermion Production during Inflation}
\label{sec.31}

We first consider fermion production during inflation. In the slow-roll regime, the background evolves sufficiently slowly that the fermion dynamics is close to adiabatic, but retains a nontrivial helicity dependence due to the derivative coupling to the inflaton. This allows for analytic control of the mode functions and makes the origin of helicity asymmetry manifest.

During slow-roll inflation, the background spacetime is well approximated by de Sitter,
\begin{equation}
a(t)=e^{H_{\inf} t},
\end{equation}
where $H_{\inf}$ is approximately constant. The conformal time is then
\begin{equation}\label{eq:conformal_time}
\tau=\int \frac{dt}{a(t)} \simeq -\frac{1}{aH_{\inf}}.
\end{equation}
We further impose the slow-roll condition
\begin{equation}
\left|\delta\right| \equiv \left|\frac{\ddot{\phi}}{H_{\inf}\dot{\phi}}\right| \ll 1,
\end{equation}
so that $\dot{\phi}$ can be treated as approximately constant. Since the fermion couples derivatively to the inflaton, only $\dot{\phi}$ enters the mode equations, and the dynamics is insensitive to the absolute value of $\phi$. Under these assumptions, the inflaton background evolves as
\begin{equation}\label{eq:inf_phi}
\phi(\tau)
=
-\frac{\dot{\phi}}{H_{\inf}}
\log\left(\frac{\tau}{\tau_{\rm in}}\right),
\end{equation}
where $\tau_{\rm in}$ denotes an initial conformal time. For the numerical estimates, we choose a benchmark slow-roll velocity corresponding to $|\dot\phi|/H_{\inf}^2=\mathcal{O}(10^3)$, which is consistent with the typical scale of large-field slow-roll inflation. Owing to the derivative coupling, 
The arbitrary reference time $\tau_{\rm in}$ only contributes an overall phase, and therefore drops out of the occupation number computed in the $\psi$ basis.
In this subsection, a dot denotes a derivative with respect to physical time.

To make the helicity structure explicit, we work in the $Y$ basis, where the derivative coupling appears as a helicity-dependent shift in the effective frequency. The equations of motion are
\begin{align}
i{\tilde{u}}_r'&=-(kr - a\dot{\theta})\tilde{u}_r + m a r\tilde{v}_r,
\\
i{\tilde{v}}_r'&=(kr - a\dot{\theta})\tilde{v}_r+  m a r\tilde{u}_r,
\end{align}
where a prime denotes a derivative with respect to conformal time.

Using $\tau=-1/(aH_\mathrm{inf})$, $H_\mathrm{inf}\approx \mathrm{const.}$, and $\dot{\theta}\approx \mathrm{const.}$, the system can be written in dimensionless form. Defining
\begin{equation}
\mu \equiv \frac{m}{H_\mathrm{inf}},
\qquad
\xi \equiv -\frac{\dot{\theta}}{H_\mathrm{inf}},
\qquad
x \equiv -\left(\frac{k}{H_{\text{inf}}}\right)(\tau H_{\text{inf}})= -k \tau.
\label{eq:mu_xi_x}
\end{equation}
With this parametrization, the equations take the form
\begin{equation}\label{approx_u_eq}
\partial_x \tilde{u}_r
=
-i\left(r+\frac{\xi}{x}\right)\tilde{u}_r
+i\frac{\mu}{x}r\tilde{v}_r,
\end{equation}
\begin{equation}\label{approx_v_eq}
\partial_x \tilde{v}_r
=
i\left(r+\frac{\xi}{x}\right)\tilde{v}_r
+i\frac{\mu}{x}r\tilde{u}_r.
\end{equation}

The parameter $\xi$ controls the helicity asymmetry through the effective shift $k \to k - r a\dot{\theta}$, while $\mu$ sets the relative importance of the mass term. In the following, we take $\xi>0$, corresponding to $\dot{\theta}<0$.

To solve these equations, it is convenient to reduce the system to a second-order differential equation. Focusing on $\tilde{v}_r(x)$ and defining
\begin{equation}
\tilde{v}_r(x)=\frac{\Phi_r(x)}{\sqrt{x}},
\end{equation}
together with the variable
\begin{equation}
z=-2ix,
\end{equation}
the mode equation can be brought into the form
\begin{equation}
\partial_z^2\Phi_r(z)+\left[
-\frac{1}{4}+\frac{\kappa_r}{z}+\frac{1}{4z^2}-\frac{\zeta^2}{z^2}
\right]\Phi_r(z)=0,
\end{equation}
which is the standard Whittaker equation with parameters
\begin{equation}
\kappa_r = r\left(\frac{1}{2}+i\xi\right),
\qquad
\zeta=i\sqrt{\mu^2 + \xi^2}.
\end{equation}
The general solution is therefore given by
\begin{equation}
\Phi_r(x) = A_r W_{\kappa_r,\zeta}(-2ix)+B_r W_{\kappa_r,\zeta}(2ix).
\end{equation}
The physical solution is selected by imposing the Bunch--Davies vacuum at early times ($x\to \infty$). Using the asymptotic behavior
\begin{equation}
W_{\kappa_r,\zeta}(2 i x)\sim e^{-ix}(2ix)^{\kappa_r}.
\end{equation}
With our convention for positive frequency in the variable $x=-k\tau$, this selects the $W_{\kappa_r,\zeta}(-2ix)$ branch.
This fixes
\begin{equation}
B_r=0.
\end{equation}
The function $\tilde{u}_r(x)$ is then determined from $\Phi_r(x)$. Defining
\begin{equation}
\tilde{u}_r(x)=\frac{\Psi_r(x)}{\sqrt{x}},
\end{equation}
one finds
\begin{equation}
\Psi_r(z)=\frac{i}{r\mu}z\left[
-\partial_z -\frac{r}{2}+\frac{1}{z}\left(
\frac{1}{2} + i\xi
\right)
\right]\Phi_r(z).
\end{equation}
Using the Whittaker identities
\begin{equation}
z \left( -\partial_z - \frac{1}{2} + \frac{\kappa}{z} \right) W_{\kappa, \zeta}(z)
=
-\left[
\left( \frac{1}{2} - \kappa \right)^2 - \zeta^2
\right] W_{\kappa-1, \zeta}(z),
\end{equation}
\begin{equation}
z \left( -\partial_z + \frac{1}{2} - \frac{\kappa}{z} \right) W_{\kappa, \zeta}(z)
=
W_{\kappa+1, \zeta}(z),
\end{equation}
the explicit solutions are obtained as
\begin{equation}
\Psi_+(z)
=
-i\mu A_+ W_{-\frac{1}{2}+i\xi,i\sqrt{\mu^2+\xi^2}}(z),
\end{equation}
\begin{equation}
\Psi_-(z)
=
-\frac{i A_-}{\mu}W_{\frac{1}{2}-i\xi,i\sqrt{\mu^2+\xi^2}}(z),
\end{equation}
where we have used
\begin{equation}
\left( \left( \frac{1}{2} - \kappa_+ \right)^2 - \zeta^2 \right)=\mu^2.
\end{equation}
Finally, imposing the normalization condition $|\tilde{u}_r|^2 + |\tilde{v}_r|^2 = 2$ fixes the overall coefficients and yields the canonically normalized mode functions,
\begin{equation}\label{u_solve}
\tilde{u}_+(x)
=
-i\mu \frac{e^{-\frac{\pi}{2}\xi}}{\sqrt{x}}
W_{-\frac{1}{2}+i\xi,i\sqrt{\mu^2+\xi^2}}(-2ix),
\qquad
\tilde{u}_-(x)
=
\frac{e^{\frac{\pi}{2}\xi}}{\sqrt{x}}
W_{\frac{1}{2}-i\xi,i\sqrt{\mu^2+\xi^2}}(-2ix),
\end{equation}
\begin{equation}\label{v_solve}
\tilde{v}_+(x)
=
\frac{e^{-\frac{\pi}{2}\xi}}{\sqrt{x}}
W_{\frac{1}{2}+i\xi,i\sqrt{\mu^2+\xi^2}}(-2ix),
\qquad
\tilde{v}_-(x)
=
i\mu \frac{e^{\frac{\pi}{2}\xi}}{\sqrt{x}}
W_{-\frac{1}{2}-i\xi,i\sqrt{\mu^2+\xi^2}}(-2ix).
\end{equation}
Substituting eqs.~\eqref{u_solve} and \eqref{v_solve} into Eq.~\eqref{eq:alpha_beta_inflation} gives the analytic Bogolyubov coefficients during inflation. For completeness, we display the relevant expression:
\begin{equation}
\begin{aligned}
\alpha_{r} &= e^{i r \varphi_{\mathbf{k}} / 2} \left[ \frac{1}{2} \sqrt{1 - \frac{kr}{\omega_k}}\, \tilde{u}_r + \frac{r}{2} \sqrt{1 + \frac{kr}{\omega_k}}\, \tilde{v}_r \right], \\
\beta_{r} &= e^{i r \varphi_{\mathbf{k}} / 2} \left[ \frac{1}{2} \sqrt{1 + \frac{kr}{\omega_k}}\, \tilde{u}_r - \frac{r}{2} \sqrt{1 - \frac{kr}{\omega_k}}\, \tilde{v}_r \right].
\end{aligned}
\end{equation}
Although the mode functions are solved in the $Y$ basis, the Bogolyubov coefficients below are those associated with the $\psi$-basis instantaneous particle number through Eq.~\eqref{eq:alpha_beta_inflation}. The occupation number is then given by
\begin{equation}\label{eq:beta_inf}
\begin{aligned}
|\beta_{r}(k,\tau)|^2
&=
\langle 0 | \hat{A}_r^\dagger(\mathbf{k},\tau) \, \hat{A}_r(\mathbf{k},\tau) | 0 \rangle \\
&= 	\frac{1}{2}+\frac{kr}{4\omega_k}\left(|\tilde{u}|^2 - |\tilde{v}|^2\right) - r \frac{ma}{2\omega_k}\mathrm{Re}\left[\tilde{u}\tilde{v}^\ast\right]\\
&= 	\frac{1}{2}+r\frac{x}{4\sqrt{x^2 + \mu^2}}\left(|\tilde{u}|^2 - |\tilde{v}|^2\right) - r \frac{\mu }{2\sqrt{x^2 + \mu^2}}\mathrm{Re}\left[\tilde{u}\tilde{v}^\ast\right].
\end{aligned}
\end{equation}
We have verified that the fermion spectrum obtained here agrees with the result of Ref.~\cite{Adshead:2018oaa}, although that analysis was carried out in the Dirac representation. This provides a nontrivial check that the physical spectrum is independent of the choice of gamma-matrix representation.

\begin{figure}[t]
    \centering
    \includegraphics[width=0.4\linewidth]{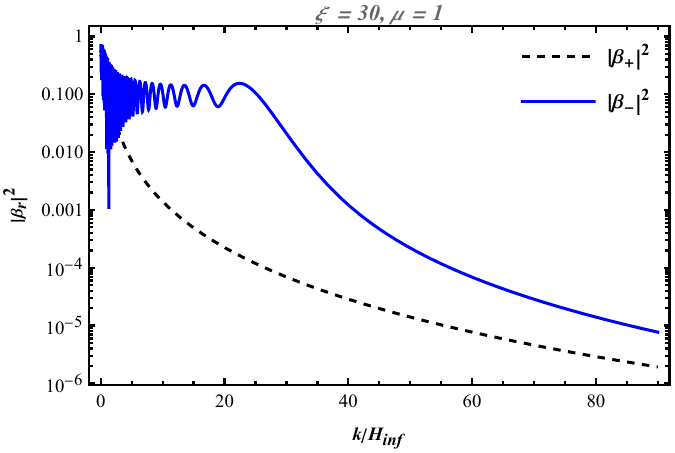}
    \includegraphics[width=0.4\linewidth]{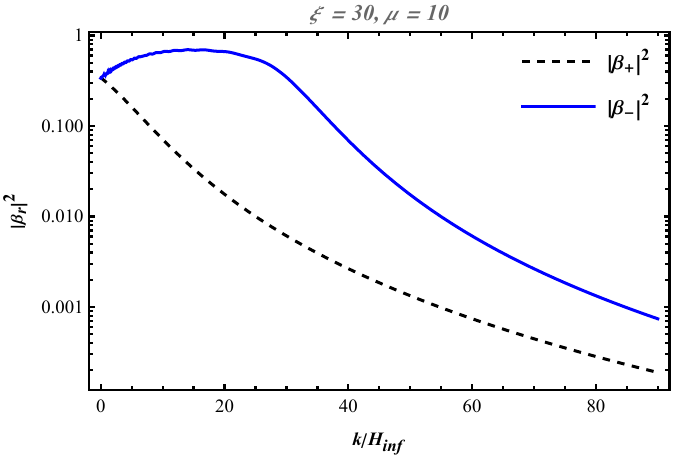}
    \includegraphics[width=0.4\linewidth]{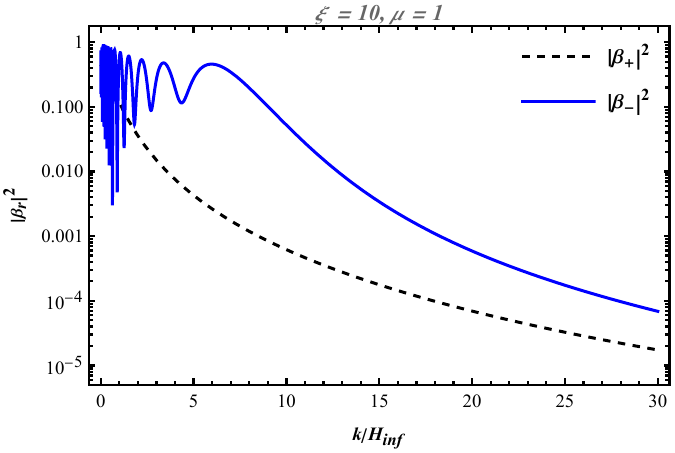}
    \includegraphics[width=0.4\linewidth]{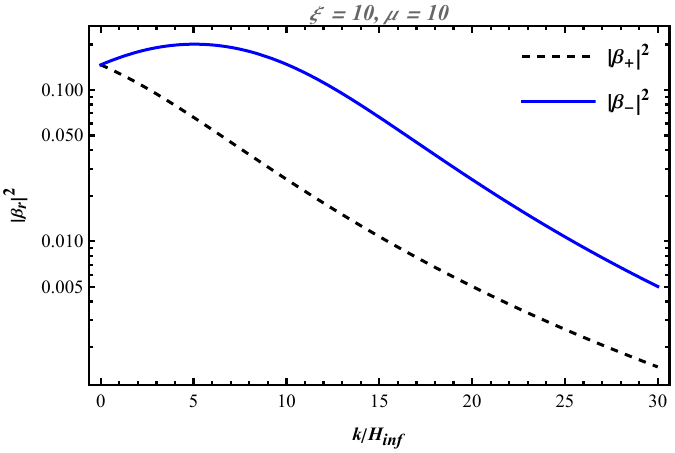}
    \caption{
    The occupation number $|\beta_r|^2$ at the end of inflation as a function of the comoving momentum $k$, for $\xi = 10,\,30$ and $\mu = 1,\,10$. Solid blue (dashed black) curves correspond to the negative-helicity $r=-1$ (positive-helicity $r=+1$ modes). For $\mu \lesssim \xi$, negative-helicity fermions are efficiently produced for $k/H_{\rm inf} \lesssim \xi$, while the positive-helicity mode remains suppressed. Helicity asymmetry is induced by the effective chemical potential of the rolling inflaton. As $\mu$ approaches or exceeds $\xi$, the asymmetry becomes less pronounced. 
    }
    \label{fig:inf_n_k}
\end{figure}

We evaluate the occupation number at the end of inflation, $a=1$ and $\tau=-1/H_{\inf}$. Fig.~\ref{fig:inf_n_k} shows $|\beta_r|^2$ as a function of the comoving momentum $k$ for representative values $\xi = 10, 30$ and $\mu = 1, 10$. The blue solid curves denote the negative-helicity mode ($r=-1$), while the black dashed curves denote the positive-helicity mode ($r=+1$). The four panels compare the dependence on the coupling strength $\xi$ and the fermion mass parameter $\mu$.

Several features emerge from the momentum dependence shown in Fig.~\ref{fig:inf_n_k}. For $\mu \lesssim \xi$, the negative-helicity mode is produced much more efficiently than the positive-helicity mode, leading to a pronounced helicity asymmetry. The results show that negative-helicity fermions are efficiently produced for $k/H_{\inf}\lesssim \xi$, or equivalently $k\lesssim |\dot{\theta}|$, while their production is rapidly suppressed for $k/H_{\inf}>\xi$. By contrast, the positive-helicity mode remains suppressed over the whole momentum range. This behavior reflects the role of the effective chemical potential induced by the rolling inflaton, which selectively enhances one helicity state. As $\mu$ approaches or exceeds $\xi$, the helicity asymmetry becomes less pronounced. In the limit $\mu \gg \xi$, fermion production is strongly suppressed for both helicities over the entire momentum range.

It is important to clarify how the inflationary contribution should be interpreted. In de Sitter space, fermion production is not localized at a single time, and the occupation number $|\beta_r|^2$ builds up continuously as the mode evolves. During inflation, the mode functions depend on $k$ and $\tau$ through the combination $x=-k\tau$. As a result, the momentum dependence shown in Fig.~\ref{fig:inf_n_k} can also be viewed as the time evolution of a fixed comoving mode after a suitable rescaling. The result evaluated at $a=1$ and $\tau=-1/H_{\inf}$ should therefore be understood as the comoving occupation number distribution accumulated during inflation.

Whether this inflationary contribution is relevant for the subsequent evolution depends on whether the produced fermions survive long enough after inflation. If the fermion lifetime is longer than about ten Hubble times around the end of inflation, for example

\begin{equation}
\Gamma_\psi^{-1}\gtrsim \mathcal{O}(10)H_{\inf}^{-1},
\end{equation}
or equivalently
\begin{equation}
\Gamma_\psi \lesssim \mathcal{O}(0.1)H_{\inf},
\end{equation}
the particles produced during inflation can persist and serve as the initial population for the subsequent preheating stage and should be included as part of the initial non-thermal population in the delayed-decay regime discussed below.

By contrast, if the fermion decays within a Hubble time after production,
\begin{equation}
\Gamma_\psi^{-1}\lesssim \mathcal{O}(H_{\inf}^{-1}),
\end{equation}
or equivalently
\begin{equation}
\Gamma_\psi \gtrsim \mathcal{O}(H_{\inf}),
\end{equation}
the inflationary contribution is efficiently removed before it can be inherited by the later fermion population. In this prompt-decay regime, we therefore do not include the inflationary contribution in the final fermion number density.

\subsection{Fermion Production from Preheating}
\label{sec.32}

After inflation, the inflaton oscillates around the minimum of its potential and drives repeated non-adiabatic evolution of the fermion modes. This stage is qualitatively different from slow-roll inflation. The inflaton background is no longer monotonic, the scale factor evolves nontrivially, and particle production occurs through a sequence of non-adiabatic events associated with the oscillating background. The final fermion abundance therefore depends not only on the production rate in each event, but also on whether the produced fermions remain in the system between successive events.

The fermion lifetime separates two physically distinct limits. If the produced fermions survive over many inflaton oscillations, the occupation number generated in earlier oscillations is carried over to later times. The same phase-space states can then remain occupied, and Pauli blocking becomes important. After sufficiently many oscillations, the spectrum approaches a saturated non-thermal distribution. By contrast, if the fermions decay rapidly after each production event, the occupation number does not accumulate from one oscillation to the next. In this case, the relevant quantity is the cumulative abundance generated by a sequence of independent production events.

We model the post-inflationary oscillating background with a quadratic potential,
\begin{equation}
V(\phi)=\frac{1}{2}m_\phi^2\phi^2,
\end{equation}
so that the inflaton and the scale factor obey~\cite{Wang:2013zva}
\begin{equation}\label{eq:phi_eom_pre}
\ddot{\phi}+3H\dot{\phi}+m_\phi^2\phi=0,
\end{equation}
and
\begin{equation}\label{eq:friedmann_pre}
3M_{\rm Pl}^2H^2=\frac{1}{2}\dot{\phi}^2+\frac{1}{2}m_\phi^2\phi^2.
\end{equation}
For the numerical analysis, we take
\begin{equation}
m_\phi=10^{13}\,{\rm GeV},
\end{equation}
and impose the initial conditions
\begin{equation}
\phi(0)= M_{\rm Pl},
\qquad
\dot{\phi}(0)=-0.7M_{\rm Pl}m_\phi,
\qquad
a(0)=1.
\end{equation}
These initial conditions give
\begin{equation}
H(0)\simeq \frac{\dot a(0)}{a(0)}\simeq 0.5\,m_\phi.
\end{equation}
Here $H(0)$ denotes the Hubble scale at the beginning of the preheating calculation, rather than the exact Hubble parameter during slow-roll inflation. Since it is of the same order as the inflationary Hubble scale, we use $H(0)\simeq H_{\inf}$ when matching the inflationary and preheating stages. The above initial conditions correspond to a time shortly after the end of inflation and determine the background evolution used below.

Throughout the numerical analysis, the comoving momentum $k$, the Hubble scale $H$, and the fermion mass $m$ are measured in units of the inflaton mass $m_\phi$. The coupling scale $f$ and the inflaton field $\phi$ are measured in units of $M_{\rm Pl}$, while time is measured in units of $m_\phi^{-1}$.

We solve the fermion dynamics in the $\psi$ basis, where the Bogolyubov coefficients and the instantaneous occupation number are defined. The $Y$ basis will still be used for physical interpretation, because in that basis the derivative coupling appears as a helicity-dependent shift of the effective momentum. Substituting the mode expansion \eqref{eq:psi_FT} into the equation of motion \eqref{eq:psi_eom}, we obtain
\begin{equation}
	\begin{aligned}
		i\dot{u}_r + r\frac{k}{a}u_r - mre^{2i\theta}v_r = 0\, , \\
		i\dot{v}_r - r\frac{k}{a}v_r - mre^{-2i\theta}u_r = 0\, .
	\end{aligned}
	\label{eq:pre_u_eq}
\end{equation}
These equations can be solved directly and then inserted into the general expressions for the Bogolyubov coefficients. For numerical evolution, however, it is more convenient to rewrite the system directly in terms of $\alpha_r$ and $\beta_r$ by using Eq.~\eqref{eq:alpha_beta_general}. This gives
\begin{equation}\label{eq:alpha_beta_eom}
\begin{aligned}
\dot{\alpha}_r
&=
-i\left(
\frac{\omega}{a}-\frac{kr}{\omega}\dot{\theta}
\right)\alpha_r
+
m\left(
\frac{kr\dot{a}}{2\omega^2}
-i\frac{a\dot{\theta}}{\omega}
\right)\beta_r,
\\ \qquad
\dot{\beta}_r
&=
-
m\left(
\frac{kr\dot{a}}{2\omega^2}
+i\frac{a\dot{\theta}}{\omega}
\right)\alpha_r
+
i\left(
\frac{\omega}{a}-\frac{kr}{\omega}\dot{\theta}
\right)\beta_r.
\end{aligned}
\end{equation}
Here
\begin{equation}
\omega=\sqrt{k^2+m^2a^2},
\end{equation}
and the time dependence of $a(t)$ and $\theta(t)$ is determined by the background solution of eqs.~\eqref{eq:phi_eom_pre} and \eqref{eq:friedmann_pre}. The occupation number $|\beta_r(k,t)|^2$ is obtained by solving Eq.~\eqref{eq:alpha_beta_eom}.

The subsequent treatment of $N_r(k,t)$ depends on the fermion lifetime. If the fermions survive over many inflaton oscillations, the same modes remain populated and Pauli blocking controls the approach to saturation. This production calculation applies to both delayed-decay regimes considered in section~\ref{sec.4}, irrespective of whether the eventual decay occurs before or after reheating. In the prompt-decay limit, the produced fermions are removed between successive non-adiabatic events, and the final abundance must be obtained by summing the contributions from individual production stages.

\subsubsection{Production in the Delayed-decay Regime}\label{sec.321}

We first consider the delayed-decay regime, in which the produced fermions do not decay immediately after each non-adiabatic event and instead survive over many inflaton oscillations. The production calculation in this subsection is therefore controlled by the condition
\begin{equation}
\Gamma_\psi\ll m_\phi .
\end{equation}
In the numerical leptogenesis estimates, we implement this hierarchy by requiring $\Gamma_\psi\lesssim m_\phi/100$.

In this regime, fermions produced at earlier times remain in the system and continue to occupy the corresponding quantum states during preheating. The occupation number therefore evolves cumulatively, and the final abundance is limited by Pauli blocking. The eventual decay may occur either before or after reheating, and this distinction will be made only when the result is applied to right-handed-neutrino leptogenesis in section~\ref{sec.4}.

For fermion production from an axion-like inflaton, the inflationary stage can provide a non-negligible initial population. Moreover, fermions produced during inflation can affect subsequent production during preheating through Pauli blocking. The Bogolyubov coefficients obtained at the end of inflation should therefore be used as the initial conditions for the preheating evolution. We numerically solve Eq.~\eqref{eq:alpha_beta_eom} with
\begin{equation}\label{eq:pre_initial_case_a}
    \alpha_r(0)= \alpha_{\mathrm{inf},r}(k,\tau_{\mathrm{end}}),
    \qquad
    \beta_r(0)= \beta_{\mathrm{inf},r}(k,\tau_{\mathrm{end}}).
\end{equation}
Here $\alpha_{\mathrm{inf},r}$ and $\beta_{\mathrm{inf},r}$ are obtained from eqs.~\eqref{eq:alpha_beta_inflation}, \eqref{u_solve}, and \eqref{v_solve}, and $\tau_{\mathrm{end}}=-1/H_{\inf}$ denotes the end of inflation.

\begin{figure}[t]
    \centering
    \includegraphics[width=0.42\linewidth]{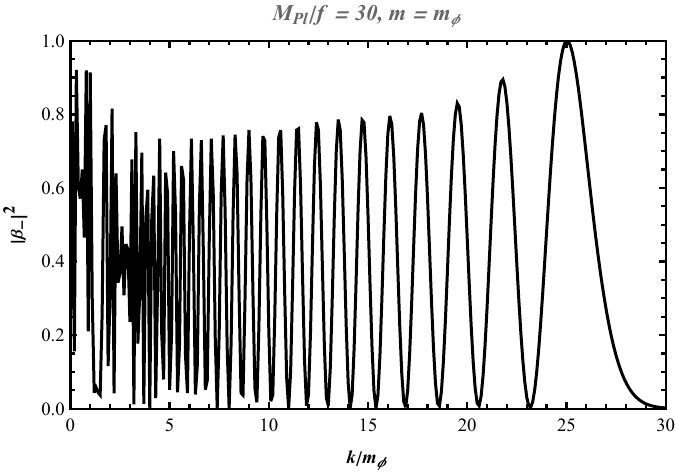}
    \includegraphics[width=0.42\linewidth]{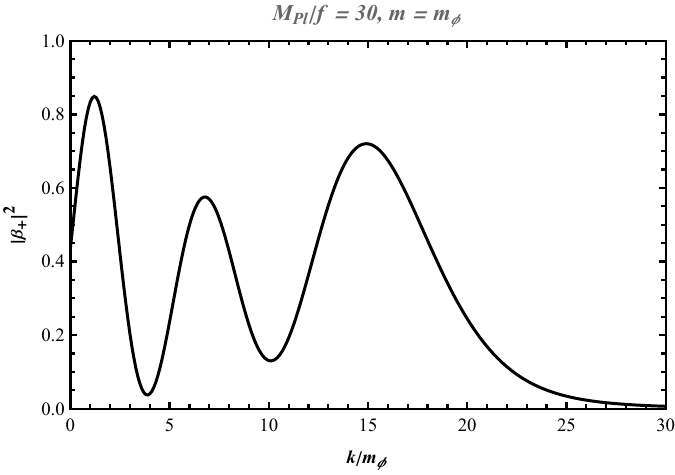}
    \includegraphics[width=0.42\linewidth]{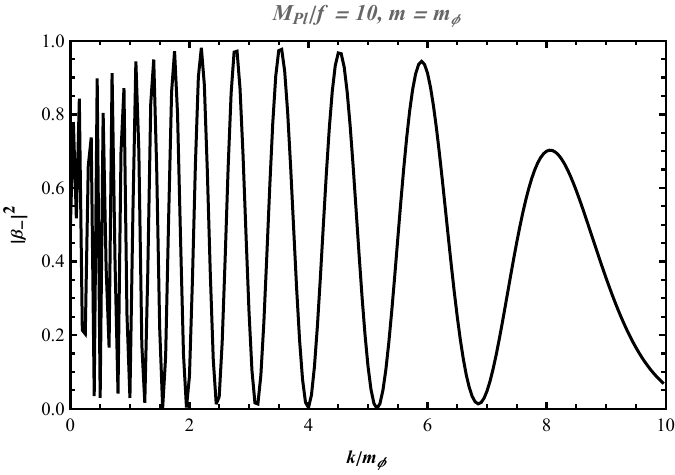}
    \includegraphics[width=0.42\linewidth]{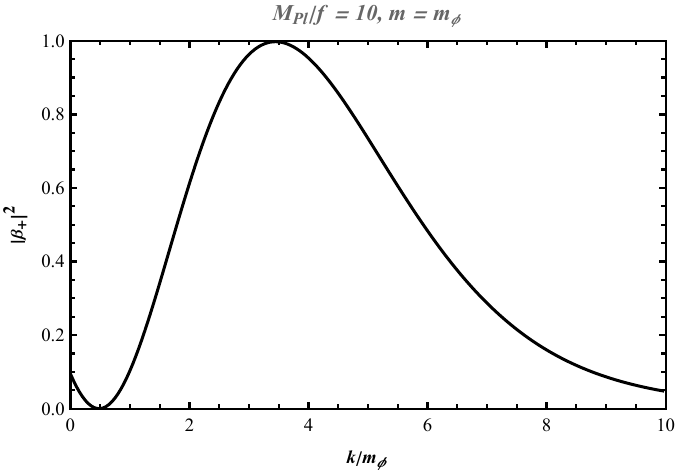}
    \caption{
    Fermion occupation number $|\beta_r(k)|^2$ during preheating as a function of the comoving momentum $k$ for delayed-decay fermions. The upper panels use $m/m_\phi=1$ and $M_{\mathrm{Pl}}/f=30$, while the lower panels use $m/m_\phi=1$ and $M_{\mathrm{Pl}}/f=10$. In each row, the left panel corresponds to the negative-helicity mode ($r=-1$), and the right panel corresponds to the positive-helicity mode ($r=+1$). The distributions are evaluated after all modes within the corresponding helicity-dependent production band have undergone their first non-adiabatic event.
    }
    \label{fig:pre_spectrum_long}
\end{figure}

We first show the occupation-number distribution $|\beta_r(k)|^2$ after the first production stage in Fig.~\ref{fig:pre_spectrum_long}. Two representative couplings, $M_{\rm Pl}/f=10$ and $30$, are considered, and the two helicities are plotted separately. As required by Pauli blocking, the occupation number always satisfies $|\beta_r|^2\leq 1$.
The distributions exhibit a finite momentum cutoff. Fermion production is efficient only for modes below this cutoff, while modes with sufficiently large $k$ are strongly suppressed. This indicates that the oscillating inflaton background does not efficiently excite arbitrarily high-momentum fermions.
The cutoff is helicity dependent. For the parameter range shown in Fig.~\ref{fig:pre_spectrum_long}, the negative-helicity branch extends to a larger maximum momentum than the positive-helicity branch,
\begin{equation}\label{eq:n_p_kmax_compare}
k_{-,{\rm max}}>k_{+,{\rm max}}.
\end{equation}
This hierarchy reflects the helicity asymmetry induced by the axion-like coupling. The derivative interaction shifts the effective momentum differently for the two helicities, so the negative-helicity branch more readily satisfies the non-adiabaticity condition and is produced over a wider momentum range.

The physical origin of fermion production during preheating is the repeated violation of adiabaticity induced by the oscillating inflaton background. This structure is less transparent in the $\psi$ basis, although this basis is convenient for defining the Bogolyubov coefficients and for numerical evolution. To make the non-adiabatic feature more explicit, it is useful to return to the $Y$ basis. Rewriting eqs.~\eqref{eq:u_eq_Y} and \eqref{eq:v_eq_Y} in cosmic time, we obtain
\begin{equation}\label{eq:Y_pre_u}
i\dot{\tilde{u}}_r+\left(\frac{rk}{a}-\dot{\theta}\right)\tilde{u}_r
=
mr\tilde{v}_r,
\end{equation}
\begin{equation}\label{eq:Y_pre_v}
i\dot{\tilde{v}}_r-\left(\frac{rk}{a}-\dot{\theta}\right)\tilde{v}_r
=
mr\tilde{u}_r.
\end{equation}
Following the approach of Refs.~\cite{Adshead:2015kza,Peloso:2000hy}, we define the helicity-dependent effective momentum and frequency,
\begin{equation}\label{eq:ktilde}
\tilde{k}_r(t)=\frac{rk}{a(t)}-\dot{\theta}(t),
\qquad
\tilde{\omega}_r(t)=\sqrt{\tilde{k}_r^2(t)+m^2}.
\end{equation}
The system becomes most non-adiabatic when $\tilde{k}_r(t)$ approaches zero. At this point, the effective frequency $\tilde{\omega}_r(t)$ reaches its minimum, and the non-adiabaticity condition
\begin{equation}
\tilde{\omega}_r^2 \ll |\dot{\tilde{\omega}}_r|
\end{equation}
is most easily satisfied. Particle production therefore occurs in bursts around the zero crossings of $\tilde{k}_r(t)$. Since $\dot{\theta}(t)$ oscillates during preheating, these zero crossings can occur repeatedly, leading to successive fermion production events.

\begin{figure}[t]
    \centering
    \includegraphics[width=0.68\linewidth]{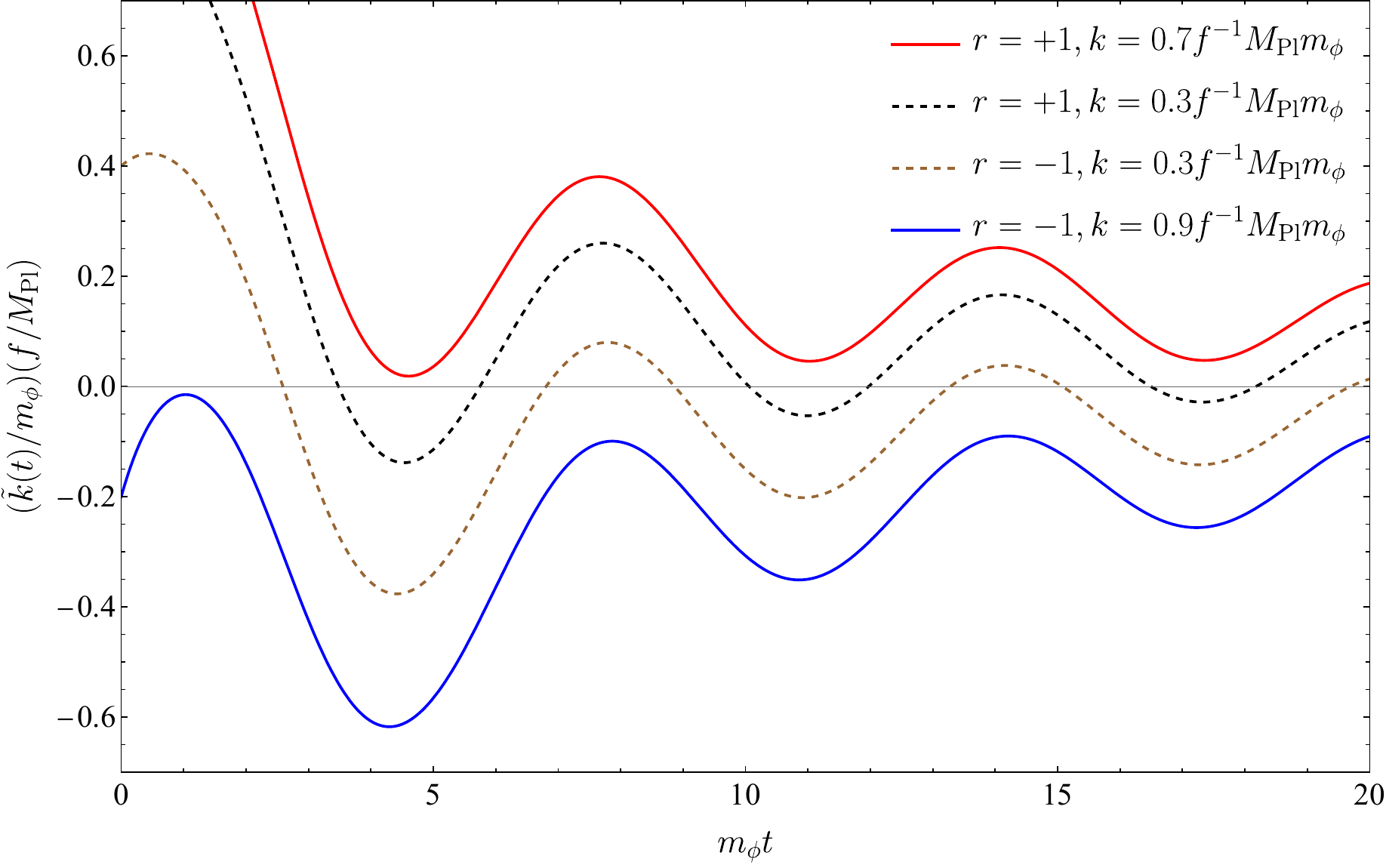}
    \caption{
    Time evolution of the helicity-dependent effective momentum $\tilde{k}_r(t)$ during preheating for representative comoving momenta. Different curves correspond to different choices of $k$ and helicity, as indicated in the plot. The parameters are the same as those used in figure~\ref{fig:pre_spectrum_long}.
    }
    \label{fig:ktilde}
\end{figure}

The momentum cutoff in Fig.~\ref{fig:pre_spectrum_long} can be understood from the time evolution of the effective momentum $\tilde{k}_r(t)$. Fig.~\ref{fig:ktilde} shows $\tilde{k}_r(t)$ for representative comoving momenta and for the two helicity branches. For small momenta, $\tilde{k}_r(t)$ crosses zero during the first few inflaton oscillations. Around these zero crossings, the adiabaticity condition is most strongly violated, so the corresponding modes can be efficiently produced.

As $k$ increases, the term $rk/a$ shifts $\tilde{k}_r(t)$ away from zero. Beyond a critical momentum, $\tilde{k}_r(t)$ no longer crosses zero, and the evolution remains adiabatic throughout the oscillation. The corresponding fermion mode is then strongly suppressed. Since the amplitude of $a\dot{\theta}$ decreases with time due to cosmic expansion, the earliest oscillations provide the largest opportunity for zero crossing. Once the earliest and largest relevant extremum of $a\dot{\theta}$ fails to generate a zero crossing, later extrema, whose amplitudes are smaller because of Hubble expansion, cannot restore efficient production.

The cutoff is also helicity dependent. This follows directly from the sign of the $rk/a$ term in $\tilde{k}_r(t)$, which shifts the two helicities in opposite directions relative to the oscillating background $\dot{\theta}(t)$. As a result, the range of momenta that can experience zero crossings differs between the two helicity branches. The cutoff momenta extracted from the numerical solutions scale approximately linearly with the coupling strength $M_{\rm Pl}/f$. In the units used in the numerical calculation, we find
\begin{equation}
\frac{k_{+,\text{max}}}{m_\phi}\simeq 0.7\,\frac{M_{\rm Pl}}{f},
\qquad
\frac{k_{-,\text{max}}}{m_\phi}\simeq 0.9\,\frac{M_{\rm Pl}}{f}.
\end{equation}
The larger value of $k_{-,\text{max}}$ is consistent with the broader negative-helicity distribution shown in Fig.~\ref{fig:pre_spectrum_long}.

\begin{figure}[htbp]
\centering
\includegraphics[width=.42\textwidth]{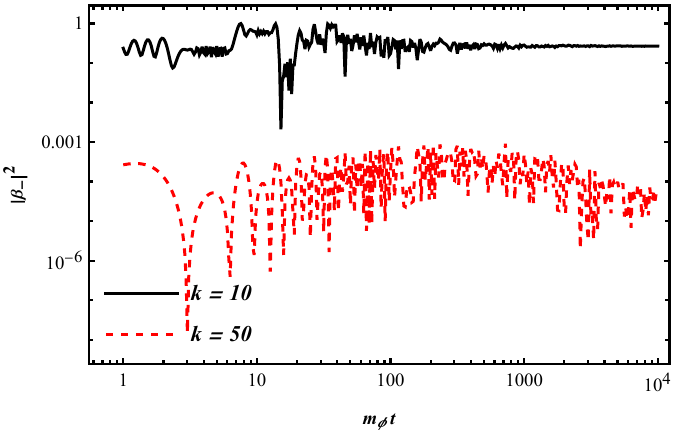}
\includegraphics[width=.42\textwidth]{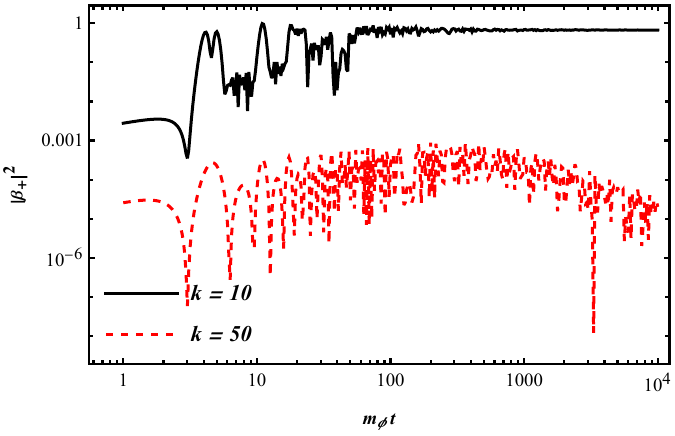}
\caption{
Time evolution of the occupation number $|\beta_r(k,t)|^2$ for representative momentum modes during preheating. The parameters are $M_{\mathrm{Pl}}/f=30$ and $m/m_\phi=1$. The left panel shows the negative-helicity mode ($r=-1$), and the right panel shows the positive-helicity mode ($r=+1$). In each panel, the curves correspond to $k=10$ and $k=50$.
}
\label{fig:beta_t}
\end{figure}

It is also useful to examine the time evolution of the occupation number for individual momentum modes. As an illustrative example, we take $M_{\mathrm{Pl}}/f=30$ and $m/m_\phi=1$. Fig.~\ref{fig:beta_t} shows $|\beta_r(k,t)|^2$ for two representative momenta, $k=10$ and $k=50$, with the two helicities shown separately. The mode with $k=10$ lies below the cutoff momentum and undergoes repeated growth during the first several non-adiabatic events. Its occupation number then approaches a saturated value, as expected from Pauli blocking.

By contrast, the mode with $k=50$ lies above the cutoff momentum. Although the corresponding Bogolyubov coefficients exhibit rapid oscillations during the background evolution, the occupation number remains strongly suppressed throughout the evolution. This confirms that high-momentum modes that do not experience zero crossings of $\tilde{k}_r(t)$ are not efficiently produced. It also justifies using $k_{r,\text{max}}$ as the effective upper limit in the momentum integration for the total fermion number density.

The total number density should in principle be obtained by integrating the full numerical occupation-number distribution. In practice, the rapid suppression of $|\beta_r(k,t)|^2$ above the helicity-dependent cutoff makes the result insensitive to the precise upper limit, as long as it is chosen above $k_{r,\text{max}}$. We therefore define the preheating contribution to the number density as
\begin{equation}
n_{r,\mathrm{pre}}(t)=\frac{1}{a^3(t)}
\int^{k_{r,\text{max}}}_{0} \frac{\text{d}^3k}{(2\pi)^3} |\beta_r(k,t)|^2,
\end{equation}
where the integral is effectively dominated by modes below the cutoff. The total preheating abundance is then
\begin{equation}
n_{\mathrm{pre}}(t)\equiv n_{+,\mathrm{pre}}(t)+n_{-,\mathrm{pre}}(t).
\end{equation}

In the delayed-decay production regime considered here, the fermions produced at earlier times remain in the system and continue to occupy the corresponding quantum states. The occupation-number distribution therefore builds up cumulatively over successive non-adiabatic events. This growth cannot continue indefinitely, because Pauli blocking limits each mode to $|\beta_r(k,t)|^2\leq 1$. As a result, the distribution gradually approaches a saturated non-thermal configuration.

Numerically, we find that the occupation numbers become approximately time independent after a characteristic time
\begin{equation}
t_{\rm stable}\sim \mathcal{O}(10^2)\,m_\phi^{-1},
\end{equation}
with only mild dependence on the model parameters within the range considered here. 
We have checked that varying the evaluation time from $200\,m_\phi^{-1}$ to $1000\,m_\phi^{-1}$ changes the final comoving number density by less than $5\%$ for the parameter range considered. The result is therefore insensitive to the precise choice of the saturation time.
We therefore evaluate the late-time preheating abundance using the saturated occupation-number distribution. For fermions with $m/m_\phi\sim 1$, the two helicities contribute at the same order of magnitude, although the negative-helicity contribution remains somewhat larger because its cutoff momentum is higher.

The qualitative behavior of the late-time abundance can be understood from the populated phase-space volume. Stronger axion--fermion coupling increases the range of momenta for which $\tilde{k}_r(t)$ crosses zero, thereby enlarging the populated region below $k_{r,\text{max}}$. It also strengthens the non-adiabaticity of each production event. At early times this leads to rapid growth of $n_{\mathrm{pre}}(t)$. Once the low-momentum modes approach saturation, however, Pauli blocking suppresses further growth. The delayed-decay regime is therefore governed by the competition between repeated non-adiabatic production and the progressive filling of the available fermionic states.

This interpretation is also relevant for incorporating the inflationary contribution into the preheating calculation. In the delayed-decay production regime, the fermions produced during inflation survive into the preheating stage and provide a helicity-asymmetric initial population. We therefore use the Bogolyubov coefficients obtained at the end of inflation as the initial conditions for the preheating evolution, rather than adding the inflationary and preheating contributions independently. The saturated abundance obtained in this way represents the final non-thermal population after the inflationary initial state has been evolved through the preheating stage, including repeated non-adiabatic production and Pauli blocking. This final saturated abundance will be used as the relevant input for the delayed-decay right-handed-neutrino scenarios.

\subsubsection{Production of Prompt-decaying Fermions}\label{sec.322}

We now consider the prompt-decay regime, in which the produced fermions decay shortly after each non-adiabatic production event. We take
\begin{equation}
\Gamma_\psi > m_\phi ,
\end{equation}
so that the fermion lifetime is shorter than the characteristic time scale $m_\phi^{-1}$ of the inflaton oscillation.
In this regime, fermions produced in one burst are removed from the system before they can persist over many inflaton oscillations. The occupation number therefore does not build up cumulatively as in the delayed-decay regime. Instead, the final abundance should be obtained by summing the contributions generated in separate production events.

The underlying production mechanism is the same as in the delayed-decay regime.
The oscillating inflaton background drives the helicity-dependent effective momentum $\tilde{k}_r(t)$ through zero, and the fermion modes undergo transient non-adiabatic evolution near these zero crossings. The difference is that the produced fermions decay before the next relevant production event. Pauli blocking from earlier bursts is therefore ineffective, and each production stage can be treated approximately as an independent event.

We implement this picture by dividing the preheating evolution into a sequence of production stages. For each stage, we compute the occupation-number distribution generated during that stage alone, starting from an effectively empty fermion state. The total abundance is then obtained by summing the number densities produced in all stages. This event-based prescription is valid when the fermion decay time is shorter than the characteristic interval between successive bursts of non-adiabatic production.

To identify the relevant production stages in the prompt-decay regime, it is useful to examine the time evolution of the background quantity $a\dot{\phi}$. Since the derivative coupling enters through
\begin{equation}
\dot{\theta}=\frac{\dot{\phi}}{f},
\end{equation}
the evolution of $a\dot{\phi}$ determines when the effective momentum $\tilde{k}_r(t)$ can cross zero for different comoving modes. In particular, the zeros and extrema of $a\dot{\phi}$ play a special role, because they determine the time intervals over which a given helicity branch can efficiently undergo non-adiabatic production.

\begin{figure}[t]
    \centering
    \includegraphics[width=0.8\linewidth]{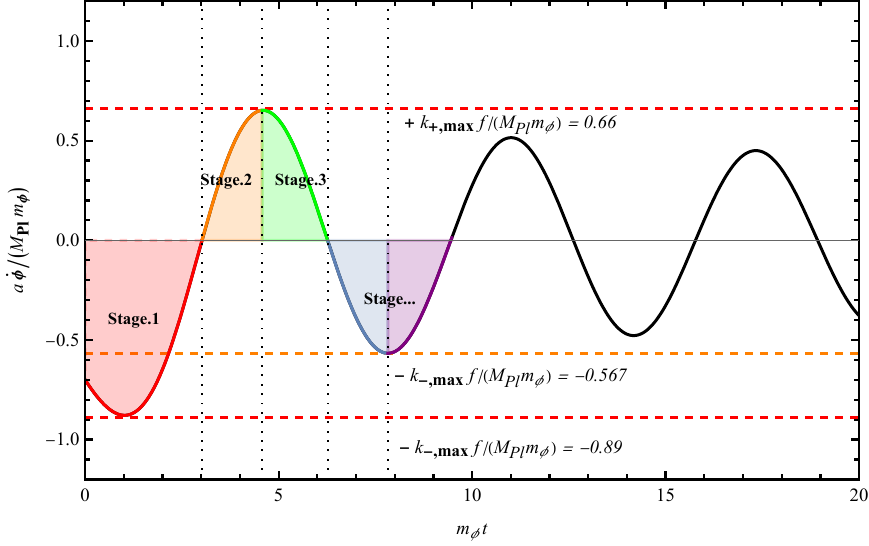}
    \caption{
    Time evolution of $a\dot{\phi}$ during preheating. The vertical dotted lines mark the zeros and extrema of $a\dot{\phi}$. The shaded regions indicate the production stages used in the event-based treatment of prompt-decaying fermions. The horizontal lines indicate the corresponding cutoff momenta $k_{r,\mathrm{max}}$ for the relevant helicity branch in each stage.
    }
    \label{fig:adotphi}
\end{figure}

Fig.~\ref{fig:adotphi} shows that $a\dot{\phi}$ oscillates with a gradually decreasing amplitude. The shaded regions in the Fig. represent the production stages used in our event-based treatment. The reason for introducing these stages is that, in the present model, different comoving modes are not produced at exactly the same time: the production condition $\tilde{k}_r(t)\simeq 0$ depends explicitly on $k$, so for a given helicity the modes with $0\leq k\leq k_{r,\mathrm{max}}$ are produced over a finite time interval rather than at a single instant. This differs from the more familiar Yukawa-coupled case, where particle production is often more sharply localized in time.

The stage structure can be read directly from Fig.~\ref{fig:adotphi}. Around each zero or extremum of $a\dot{\phi}$, one of the two helicity branches can experience non-adiabatic production over a range of comoving momenta. We therefore define a production stage as the time interval during which all modes in the relevant momentum range for a given helicity have undergone one production event. In Fig.~\ref{fig:adotphi}, each shaded region corresponds to one such stage. Concretely, the first shaded region represents the first production stage of one helicity branch; the next shaded region corresponds to the first production stage of the opposite helicity branch; the following shaded regions then describe the subsequent stages in alternating order. In this way, Fig.~\ref{fig:adotphi} provides a convenient map of the full sequence of production events during preheating.

Fig.~\ref{fig:adotphi} also makes clear why the stages are bounded by the zeros and extrema of $a\dot{\phi}$. Within a given shaded interval, all modes with $0\leq k\leq k_{r,\mathrm{max}}$ for the relevant helicity can satisfy $\tilde{k}_r(t)\simeq 0$ once. The end of that shaded interval is then defined as the time by which the full momentum range of that helicity has completed one production event. The horizontal lines shown in Fig.~\ref{fig:adotphi} indicate the corresponding cutoff momenta $k_{r,\mathrm{max}}$ for each stage.

Because the amplitude of $a\dot{\phi}$ decreases with time, the range of producible momenta also shrinks from one stage to the next, as can be seen directly from the progressively smaller shaded regions in Fig.~\ref{fig:adotphi}. Production eventually shuts off altogether when the oscillation amplitude becomes too small to overcome the fermion mass. A simple criterion is obtained by comparing the maximal value of $|\dot{\theta}|$ in a given burst with the fermion mass $m$: once
\begin{equation}\label{eq:stop_condition}
|\dot{\theta}| < m,
\end{equation}
the effective frequency $\tilde{\omega}_r=\sqrt{\tilde{k}_r^2+m^2}$ is dominated by the mass term, the evolution becomes adiabatic, and no substantial further production occurs. We therefore retain only those extrema for which $|\dot{\theta}|\gtrsim m$, together with the zeros between them, as physically relevant production stages.

Let $t_n$ denote the time at which the $n$th production stage ends. Each stage, including the first one, is evolved from the vacuum initial state, and the evolution is terminated at the end of that stage. For example,
the first stage is computed in the same way as in the delayed-decay regime, except that the initial condition is vacuum initial condition, and the evolution is terminated at $t=t_1$. For later stages, we assume that the fermions produced in the previous stage have already decayed before the next stage begins. Under this assumption, the produced fermion states are effectively emptied between successive production events. We further assume that the decay products do not significantly back-react on the inflaton background and do not repopulate the fermion states relevant for the next burst. Therefore, for each production stage we reset the Bogolyubov coefficients to the vacuum initial condition,
\begin{equation}\label{eq:initial_fast_decay}
\alpha_r(0)=1,
\qquad
\beta_r(0)=0.
\end{equation}

The $n$th stage is then evolved over the interval
\begin{equation}
t\in [0,\, t_n-t_{n-1}],
\end{equation}
with the background quantities evaluated at the shifted time $t+t_{n-1}$. In this way, the local time variable $t$ describes the evolution within a single production stage, while the physical background follows the original preheating solution. The Bogolyubov-coefficient equations for the $n$th stage are therefore
\begin{equation}\label{eq:fast_decay_stage}
\begin{aligned}
\dot{\alpha}_r
&=
-i\left(
\frac{\omega}{a(t+t_{n-1})}-\frac{kr}{\omega}\dot{\theta}(t+t_{n-1})
\right)\alpha_r
\\
&\quad
+
m\left(
\frac{kr\dot{a}(t+t_{n-1})}{2\omega^2}
-i\frac{a(t+t_{n-1})\dot{\theta}(t+t_{n-1})}{\omega}
\right)\beta_r,
\\
\dot{\beta}_r
&=
-
m\left(
\frac{kr\dot{a}(t+t_{n-1})}{2\omega^2}
+i\frac{a(t+t_{n-1})\dot{\theta}(t+t_{n-1})}{\omega}
\right)\alpha_r
\\
&\quad
+
i\left(
\frac{\omega}{a(t+t_{n-1})}-\frac{kr}{\omega}\dot{\theta}(t+t_{n-1})
\right)\beta_r.
\end{aligned}
\end{equation}
This prescription isolates the fermion spectrum generated in each production stage and avoids double counting between successive bursts.

Since the maximal producible momentum changes from one stage to another, the momentum integral must be truncated separately for each event. A natural estimate of the stage-dependent cutoff $k_{r,\mathrm{max},n}$ is obtained from the extremum of $a\dot{\phi}$ associated with that burst, using Eq.~\eqref{eq:ktilde}. Denoting this time by $t_{\rm ext}$, the characteristic cutoff scale is
\begin{equation}
k_{r,\mathrm{max},n}\simeq a(t_{\rm ext})|\dot{\theta}(t_{\rm ext})|=a(t_{\rm ext})\frac{|\dot{\phi}(t_{\rm ext})|}{f}.
\end{equation}
Numerically, however, modes slightly above this estimate can still be excited because the non-adiabatic region has a finite temporal width. We therefore use
\begin{equation}
k_{r,\mathrm{max},n}
=
C\,a(t_{\rm ext})|\dot{\theta}(t_{\rm ext})|,
\end{equation}
with
\begin{equation}
C\simeq 1.1.
\end{equation}
The factor $C$ is introduced only to include the small spectral tail near the cutoff. It should be regarded as a practical numerical choice rather than a fundamental physical constant. We have checked that varying $C$ from $1.0$ to $1.2$ changes the total abundance only mildly, indicating that the result is insensitive to the precise treatment of the spectral tail.

\begin{figure}[t]
    \centering
    \includegraphics[width=0.42\linewidth]{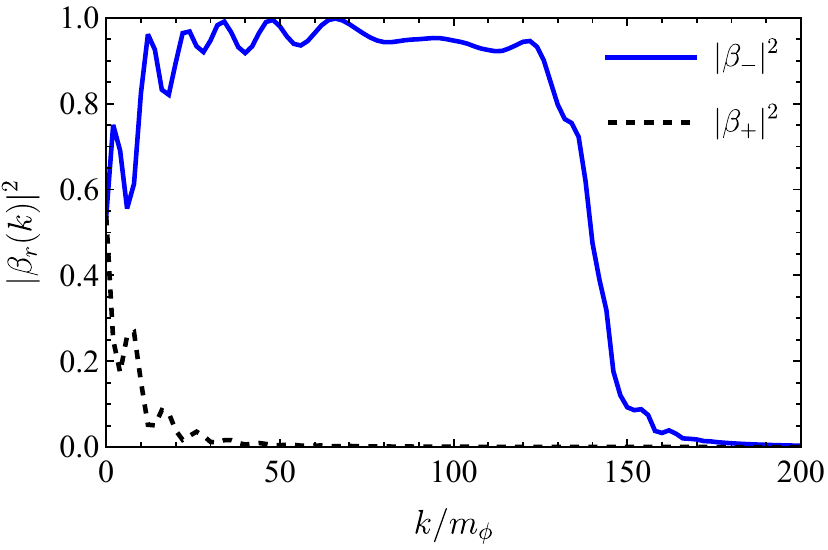}
    \includegraphics[width=0.42\linewidth]{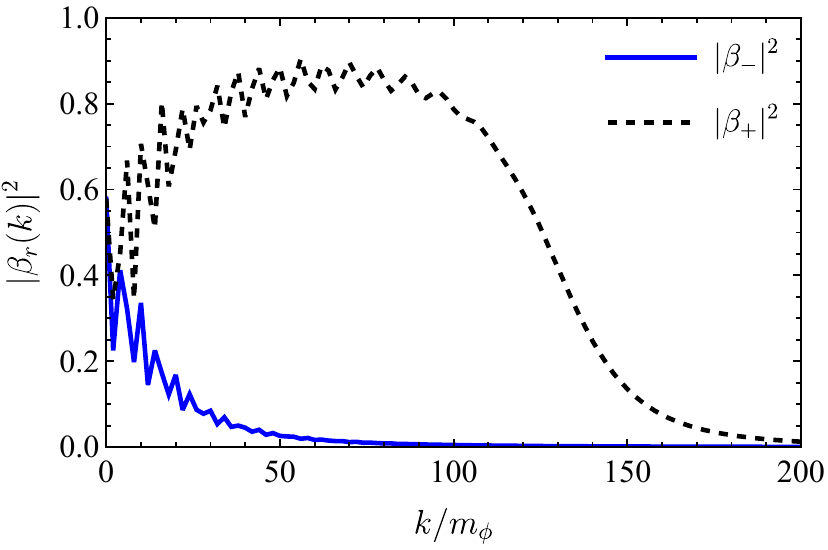}
    \includegraphics[width=0.42\linewidth]{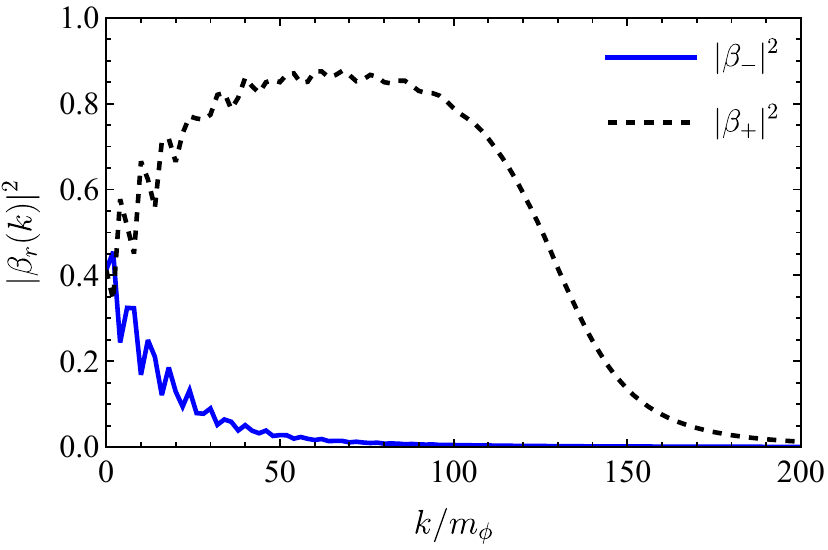}
    \includegraphics[width=0.42\linewidth]{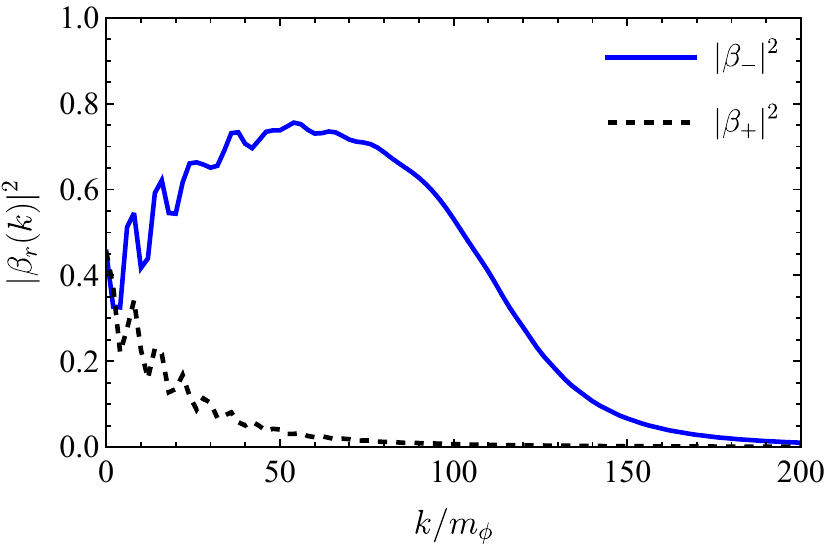}
    \caption{
    Fermion occupation-number distributions produced in the first four production stages in the prompt-decay regime. The parameters are $M_{\mathrm{Pl}}/f=200$ and $m=10\,m_\phi=10^{14}\,\mathrm{GeV}$. Each panel shows the distribution generated during one production stage, with the two helicities plotted separately.
    }
    \label{fig:fast_decay_spectrum}
\end{figure}

Using this event-based prescription, we compute the occupation-number distributions generated in the first several production stages. Representative results are shown in Fig.~\ref{fig:fast_decay_spectrum}, where we take $M_{\mathrm{Pl}}/f=200$ and $m=10\,m_\phi=10^{14}\,\mathrm{GeV}$. The four panels correspond to the first four production stages. Successive stages preferentially produce opposite helicities, following the alternating sign of the oscillating inflaton velocity. The occupation number generated in later stages is smaller than that in earlier stages, because the amplitude of the inflaton oscillation decreases with time and the non-adiabaticity condition becomes harder to satisfy.

To characterize the cumulative abundance in the prompt-decay regime, we define the comoving number density contributed by the first $N$ production stages as
\begin{equation}
N_{\psi}^{(N)}
=
\sum_{n=1}^{N}\Delta N_n,
\end{equation}
where $\Delta N_n$ denotes the comoving number density produced in the $n$th stage. Each contribution is computed from the occupation-number distribution generated during that stage alone, with the stage-dependent momentum cutoff discussed above. This definition implements the physical assumption that the fermions produced in one stage decay before the next stage begins, so that the final abundance is obtained by summing the yields from successive production events.

\begin{figure}[t]
    \centering
    \includegraphics[width=0.62\linewidth]{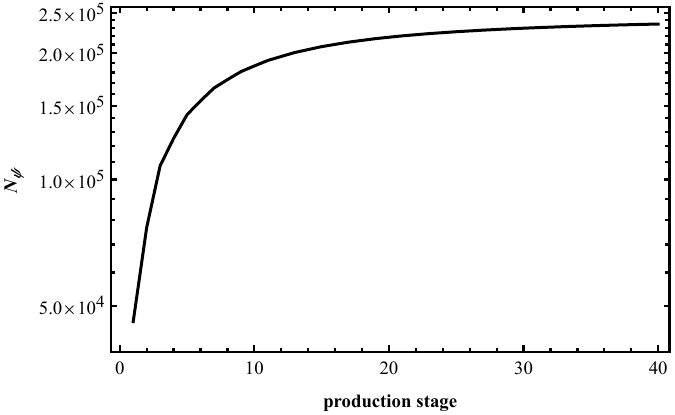}
    \caption{
    Cumulative comoving fermion number density as a function of the number of production stages in the prompt-decay regime. The parameters are $M_{\mathrm{Pl}}/f=200$ and $m=10\,m_\phi=10^{14}\,\mathrm{GeV}$. The curve is obtained by summing the stage-by-stage comoving yields.
    }
    \label{fig:fast_decay_cumulative}
\end{figure}

The resulting cumulative production is shown in Fig.~\ref{fig:fast_decay_cumulative}, where we take $M_{\mathrm{Pl}}/f=200$ and $m=10\,m_\phi=10^{14}\,\mathrm{GeV}$. The abundance grows rapidly during the first few stages and then gradually approaches a plateau. This behavior reflects the decreasing efficiency of particle production as the inflaton oscillation amplitude redshifts away. In the representative example shown in Fig.~\ref{fig:fast_decay_cumulative}, the first 20 production stages provide the dominant non-negligible contributions. After the 20th stage, the cumulative comoving number density changes only mildly, indicating that later bursts are already inefficient.

This cumulative result also clarifies the difference between the prompt-decay and delayed-decay regimes. For the same background evolution and coupling parameters, the prompt-decay regime can yield a larger total fermion abundance than the delayed-decay regime. The reason is not that a single production event is necessarily more efficient. Rather, the difference comes from how successive events contribute to the final abundance.

In the delayed-decay regime, fermions produced at earlier times remain in the system and continue to occupy the relevant momentum states. Once the occupation number becomes sizable, Pauli blocking suppresses further production in the same states. In the prompt-decay regime, by contrast, the particles produced in one stage are removed before the next stage begins. The same momentum region can therefore be populated repeatedly in successive bursts. Equivalently, the suppression due to persistent final-state occupation is strongly reduced, and the total abundance receives approximately additive contributions from multiple production stages.

These results will be used in section~4 when discussing baryogenesis from heavy right-handed neutrinos. In particular, the prompt-decay regime provides a natural framework in which the produced fermions can decay promptly into lepton-number-violating final states, thereby converting event-by-event fermion production during preheating into a net baryon asymmetry after sphaleron processing.

Before turning to leptogenesis, let us summarize the production history. Inflation generates a helicity-asymmetric initial spectrum whose size depends on the fermion mass and the axion--fermion coupling. Preheating then drives repeated non-adiabatic production through the oscillating inflaton background. In the delayed-decay regime, the inflationary population is evolved through preheating and the final abundance is controlled by the saturated occupation-number distribution. In the prompt-decay regime, the relevant quantity is instead the sum of the stage-by-stage yields. This distinction is the main reason why the baryogenesis analysis in the next section must treat the two regimes separately, even though both originate from the same derivative interaction.

\section{Baryogenesis from Non-thermally Produced Right-handed Neutrinos}
\label{sec.4}

In this section, we apply the fermion-production results obtained in section~3 to non-thermal leptogenesis. The basic idea is that the axion-like inflaton can non-thermally produce heavy Majorana fermions during inflation and preheating. If these fermions are identified with right-handed neutrinos, their subsequent CP-violating decays can generate a lepton asymmetry, which is then partially converted into a baryon asymmetry by electroweak sphaleron processes.

Our purpose is not to construct a complete thermal leptogenesis scenario, but to estimate whether the non-thermal abundance generated by the inflaton can provide a viable source for the observed baryon asymmetry. Motivated by the type-I seesaw mechanism, we identify the produced Majorana fermion with the lightest right-handed neutrino $N_1$. The production history derived in section~3 determines the non-thermal abundance of $N_1$, while the standard leptogenesis parameters determine how efficiently this abundance is converted into a lepton asymmetry.

A central point is that the two preheating regimes discussed in section~3 lead to different leptogenesis estimates. In the delayed-decay production regime, the heavy neutrinos survive over many inflaton oscillations. The relevant input is therefore the final saturated non-thermal abundance obtained after the inflationary initial population has been evolved through preheating, including Pauli blocking. In the prompt-decay regime, the heavy neutrinos decay shortly after each production burst. The relevant input is instead the sum of the stage-by-stage yields. Keeping this distinction explicit is essential, because the abundance entering the baryon-asymmetry estimate is defined differently in the two regimes.

We consider a standard leptogenesis setup with three generations of right-handed neutrinos,
\begin{equation}
N_{Ri}\sim (1,0)_{SU(2)_L\times U(1)_Y},
\end{equation}
whose interactions are described by
\begin{equation}\label{eq:NR_lag}
\mathcal{L}
=
\overline{N_R} i \slashed{\partial} N_R
-
\left(
\frac{1}{2} \overline{N_R^c} M N_R
+
\overline{\ell_L} Y_\nu \widetilde{H} N_R
+
{\rm h.c.}
\right),
\end{equation}
where $\widetilde{H}=i\sigma_2 H^\ast$, $\ell_L$ is the left-handed lepton doublet, and $M$ is the Majorana mass matrix of the heavy neutrinos. We work in the basis in which
\begin{equation}
M={\rm diag}(M_1,M_2,M_3),
\qquad
M_1<M_2<M_3.
\end{equation}
The light-neutrino mass matrix is then generated by the seesaw relation
\begin{equation}
\label{light neu mass}
m_\nu = -m_D M^{-1} m_D^T,
\qquad
m_D=\frac{Y_\nu v}{\sqrt{2}},
\end{equation}
with $v=247~{\rm GeV}$.

The lepton asymmetry is generated by the CP-violating decay of the lightest right-handed neutrino $N_1$,
\begin{equation}
N_1 \rightarrow \ell + H,
\qquad
N_1 \rightarrow \bar{\ell} + H^\ast.
\end{equation}
The CP asymmetry parameter is defined as
\begin{equation}\label{eq:cpasym}
\epsilon
=
\frac{
\Gamma_{N_1}(N_1\to \ell+H)
-
\Gamma_{N_1}(N_1\to \bar{\ell}+H^\ast)
}{
\Gamma_{N_1}(N_1\to \ell+H)
+
\Gamma_{N_1}(N_1\to \bar{\ell}+H^\ast)
}.
\end{equation}
For the purposes of our estimate, the total CP asymmetry receives the usual vertex and self-energy contributions,
\begin{equation}
\epsilon=\epsilon_V+\epsilon_S.
\end{equation}
For completeness, we quote the standard expressions~\cite{Luty:1992un,Plumacher:1996kc,Giudice:1999fb},
\begin{equation}
\epsilon_V
=
\frac{1}{8\pi (Y_\nu^\dagger Y_\nu)_{11}}
\sum_{j=2,3}
{\rm Im}\!\left[(Y_\nu^\dagger Y_\nu)_{1j}^2\right]
F\!\left(\frac{M_j^2}{M_1^2}\right),
\end{equation}
with
\begin{equation}
F(x)=\sqrt{x}\left[1-(1+x)\ln\left(\frac{1+x}{x}\right)\right],
\end{equation}
and, in the resonant self-energy form~\cite{Giudice:1999fb,Pilaftsis:1998pd},
\begin{equation}
\epsilon_S
=
\frac{
{\rm Im}\!\left[(Y_\nu^\dagger Y_\nu)_{12}^2\right]
}{
(Y_\nu^\dagger Y_\nu)_{11}(Y_\nu^\dagger Y_\nu)_{22}
}
\left[
\frac{(M_1^2-M_2^2)M_1\Gamma_{N_2}}
{(M_1^2-M_2^2)^2+M_1^2\Gamma_{N_2}^2}
\right].
\end{equation}
The total tree-level decay width of $N_i$ is
\begin{equation}\label{eq:GammaNi}
\Gamma_{N_i}
=
\frac{(Y_\nu^\dagger Y_\nu)_{ii}}{8\pi}M_i.
\end{equation}
It is convenient to introduce the effective neutrino mass
\begin{equation}\label{eq:mtilde}
\widetilde{m}_1
=
\frac{(m_D^\dagger m_D)_{11}}{M_1}
=
\frac{(Y_\nu^\dagger Y_\nu)_{11}v^2}{2M_1},
\end{equation}
in terms of which the decay width of $N_1$ can be written as
\begin{equation}\label{eq:GammaN1_mtilde}
\Gamma_{N_1}
=
\frac{\widetilde{m}_1 M_1^2}{4\pi v^2}.
\end{equation}

We now examine whether the right-handed neutrinos produced non-thermally during preheating can generate a sufficiently large baryon asymmetry through their subsequent decays. The resulting asymmetry is controlled by the interplay among three characteristic time scales: the inflaton oscillation time scale $m_\phi^{-1}$, the inflaton decay time scale $\Gamma_\phi^{-1}$, and the heavy-neutrino decay time scale $\Gamma_{N_1}^{-1}$. In addition, possible washout processes after the completion of reheating may further suppress the generated lepton asymmetry. According to the relative size of $\Gamma_{N_1}$, $\Gamma_\phi$, and $m_\phi$, the evolution of the produced $N_1$ population can be divided into three physically distinct regimes:

\begin{enumerate}
    \item \textbf{Pre-reheating delayed decay:} $\Gamma_\phi<\Gamma_{N_1}\ll m_\phi$.  
    In this regime, $N_1$ lives much longer than one inflaton oscillation period, so its non-thermal abundance can build up over many oscillations and eventually saturate due to Pauli blocking. Nevertheless, it decays before the inflaton decay completes, namely before reheating occurs.
    \label{case_a}

    \item \textbf{Post-reheating delayed decay:} $\Gamma_{N_1}<\Gamma_\phi$.  
    In this regime, the produced $N_1$ particles remain present until the inflaton has completed its decay and subsequently decay in the radiation-dominated era.
    \label{case_b}

    \item \textbf{Prompt decay after each production burst:} $\Gamma_{N_1}>m_\phi$.  
    In this regime, $N_1$ decays within a single inflaton oscillation period. Each preheating burst still produces a nonzero $N_1$ abundance, but this abundance decays before the next burst becomes relevant. Therefore, successive bursts contribute independently to the final asymmetry, without forming a saturated $N_1$ population across many oscillations.
    \label{case_c}
\end{enumerate}

In the following analysis, we focus on the genuinely non-thermal regime in which the abundance of $N_1$ is dominated by particle production before the onset of the radiation-dominated era, including both inflationary production and preheating production. We therefore require that $N_1$ is not efficiently thermalized after reheating is completed. A simple and conservative way to ensure this is to impose
\begin{equation}
M_1>T_{\rm RH}.
\end{equation}
If instead $T_{\rm RH}\gtrsim M_1$, the thermal bath produced by inflaton decay can efficiently regenerate $N_1$. In that case, the subsequent generation of the lepton asymmetry would be largely governed by the standard thermal leptogenesis dynamics rather than by the non-thermal abundance produced during inflation and preheating. Residual inverse decays and off-shell scatterings for $T_{\rm RH}<M_1$ depend on the Yukawa couplings and would require a Boltzmann or density-matrix treatment; we do not include these effects in the present estimate.

We further require
\begin{equation}
M_1>\frac{m_\phi}{2},
\end{equation}
so that the perturbative decay channel $\phi\to N_1N_1$ is kinematically forbidden. If $M_1<m_\phi/2$, the same inflaton--fermion interaction responsible for the non-perturbative production can also induce perturbative inflaton decays into $N_1N_1$. This would provide an additional source of right-handed neutrinos and an irreducible contribution to the total inflaton width, so that the lifetime classification based on an externally fixed $\Gamma_\phi$ would no longer apply without modification. Since this regime is more model dependent and involves the standard inflaton-decay contribution to non-thermal leptogenesis, we restrict our analysis to $M_1>m_\phi/2$.

We adopt the instantaneous-reheating approximation for the inflaton decay. The reheating temperature $T_{\rm RH}$ is defined by equating the total energy density at the completion of inflaton decay to the radiation energy density,
\begin{equation}\label{eq:reheat_rho}
3M_{\rm Pl}^2 H^2(T_{\rm RH})
=
\frac{\pi^2}{30}g_\ast T_{\rm RH}^4 .
\end{equation}
In this approximation, reheating occurs when the Hubble rate becomes comparable to the inflaton decay rate,
\begin{equation}
H(T_{\rm RH})=\Gamma_\phi .
\end{equation}
Combining the above two relations gives
\begin{equation}\label{eq:Gamma_phi_TRH}
\Gamma_\phi
=
\sqrt{\frac{\pi^2 g_\ast}{90}}\,
\frac{T_{\rm RH}^2}{M_{\rm Pl}} ,
\end{equation}
where throughout this section we take $g_\ast=106.75$.
These relations allow us to express the lifetime conditions of $N_1$ in terms of the reheating temperature and, equivalently, to translate them into constraints on the seesaw parameters $(M_1,\widetilde m_1)$. Together with the non-thermal condition $M_1>T_{\rm RH}$, they define the parameter region in which the final baryon asymmetry is controlled primarily by the non-thermally produced $N_1$ population.

There is one additional constraint. After reheating is completed, the thermal bath can induce lepton-number-violating $\Delta L=2$ scattering processes. If these processes remain in equilibrium, they can wash out the lepton asymmetry generated from the decays of the non-thermally produced $N_1$ particles. Therefore, the reheating temperature cannot be arbitrarily high. More explicitly, in the radiation-dominated era after reheating, the generated lepton asymmetry can be erased by the processes
\begin{equation}
\ell H \leftrightarrow \bar{\ell} H^\ast,
\qquad
\ell\ell \leftrightarrow H^\ast H^\ast ,
\end{equation}
which are mediated by the virtual exchange of heavy right-handed neutrinos. These processes can be active even if the $N_1$ particles themselves have already decayed or have never been thermalized. To preserve the generated lepton asymmetry, the corresponding washout rate should be smaller than the Hubble expansion rate at the reheating temperature,
\begin{equation}
\label{avoid washout condition}
    \Gamma_{\Delta L=2}(T_{\rm RH}) < H(T_{\rm RH}) .
\end{equation}

For $T<M_1$, the heavy right-handed neutrinos can be integrated out, and the $\Delta L=2$ scattering processes are described by the Weinberg operator. In this limit, the squared scattering amplitude can be approximated as~\cite{Buchmuller:1997yu,Plumacher:1996kc,Buchmuller:2004nz,Garbrecht:2024xfs}
\begin{equation}
|\mathcal{M}_{\Delta L=2}(s)|^2
=
\frac{48\,s}{v^4}
\mathrm{Tr}\left(m_\nu^\dagger m_\nu\right) ,
\end{equation}
where $m_\nu$ is the light-neutrino mass matrix given in Eq.~\eqref{light neu mass}, and $s$ denotes the center-of-mass energy squared.
For massless initial- and final-state particles, the corresponding scattering cross section is
\begin{equation}
    \sigma_{\Delta L=2}(s)
    =
    \frac{1}{16\pi s^2}
    \int_{-s}^{0}\text{d}u\,
    |\mathcal{M}_{\Delta L=2}(s)|^2
    =
    \frac{3}{\pi v^4}
    \mathrm{Tr}\left(m_\nu^\dagger m_\nu\right) ,
\end{equation}
where $u$ is the Mandelstam variable.

The thermally averaged reaction density is given by~\cite{Giudice:2003jh}
\begin{equation}
\gamma_{\Delta L=2}(T)
=
\frac{T}{64\pi^4}
\int_{0}^{\infty}\text{d}s\,
\sqrt{s}\,
\hat{\sigma}_{\Delta L=2}(s)\,
K_1\left(\frac{\sqrt{s}}{T}\right) ,
\end{equation}
where $K_1$ is the modified Bessel function of the second kind, and $\hat{\sigma}_{\Delta L=2}(s)$ is the reduced cross section. For massless particles, one has
\begin{equation}
\hat{\sigma}_{\Delta L=2}(s)=2s\,\sigma_{\Delta L=2}(s).
\end{equation}
Using the identity
\begin{equation}
\int_0^\infty \text{d}x\,x^{\mu-1}K_\nu(x)
=
2^{\mu-2}
\Gamma\left(\frac{\mu+\nu}{2}\right)
\Gamma\left(\frac{\mu-\nu}{2}\right),
\end{equation}
we obtain
\begin{equation}
    \gamma_{\Delta L=2}(T)
    =
    \frac{3T^6}{\pi^5 v^4}
    \mathrm{Tr}\left(m_\nu^\dagger m_\nu\right) .
\end{equation}
The corresponding washout rate per lepton is then
\begin{equation}
\label{washout reaction rate}
    \Gamma_{\Delta L=2}(T)
    =
    \frac{\gamma_{\Delta L=2}(T)}
    {n_\ell^{\rm eq}(T)}
    =
    \frac{2T^3}
    {\pi^3 v^4\zeta(3)}
    \mathrm{Tr}\left(m_\nu^\dagger m_\nu\right) ,
\end{equation}
where we have used the equilibrium number density for one relativistic lepton doublet degree of freedom,
\begin{equation}
    n_\ell^{\rm eq}(T)
    =
    \frac{3\zeta(3)}{2\pi^2}T^3 .
\end{equation}
The trace is basis independent and can be written in terms of the light-neutrino mass eigenvalues as
\begin{equation}
    \mathrm{Tr}\left(m_\nu^\dagger m_\nu\right)
    =
    m_{\nu_1}^2+m_{\nu_2}^2+m_{\nu_3}^2 .
\end{equation}
For the numerical estimate below, we take the normal-ordering benchmark with the lightest neutrino mass much smaller than the measured mass splittings, which gives
\begin{equation}\label{eq:summ2bound}
    m_{\nu_1}^2+m_{\nu_2}^2+m_{\nu_3}^2
    \simeq
    2.57\times10^{-3}\,{\rm eV}^2 .
\end{equation}
Combining Eqs.~\eqref{avoid washout condition}, \eqref{eq:Gamma_phi_TRH}, and \eqref{washout reaction rate}, the condition that the $\Delta L=2$ scattering processes are out of equilibrium at reheating gives an upper bound on the reheating temperature,
\begin{equation}
    T_{\rm RH}
    <
    \frac{\pi^3\zeta(3)v^4}{2\,\mathrm{Tr}(m_\nu^\dagger m_\nu)}
    \sqrt{\frac{\pi^2g_\ast}{90}}\,
    \frac{1}{M_{\rm Pl}} .
\end{equation}
Taking $g_\ast=106.75$, $v=247\,{\rm GeV}$, and the benchmark value in Eq.~\eqref{eq:summ2bound}, we find
\begin{equation}
    T_{\rm RH}
    <
    4\times10^{13}\,{\rm GeV}.
\end{equation}
This bound ensures that the lepton asymmetry generated from the non-thermally produced $N_1$ population is not subsequently erased by $\Delta L=2$ washout after reheating.

\begin{figure}[t]
\centering
\includegraphics[width=0.62\linewidth]{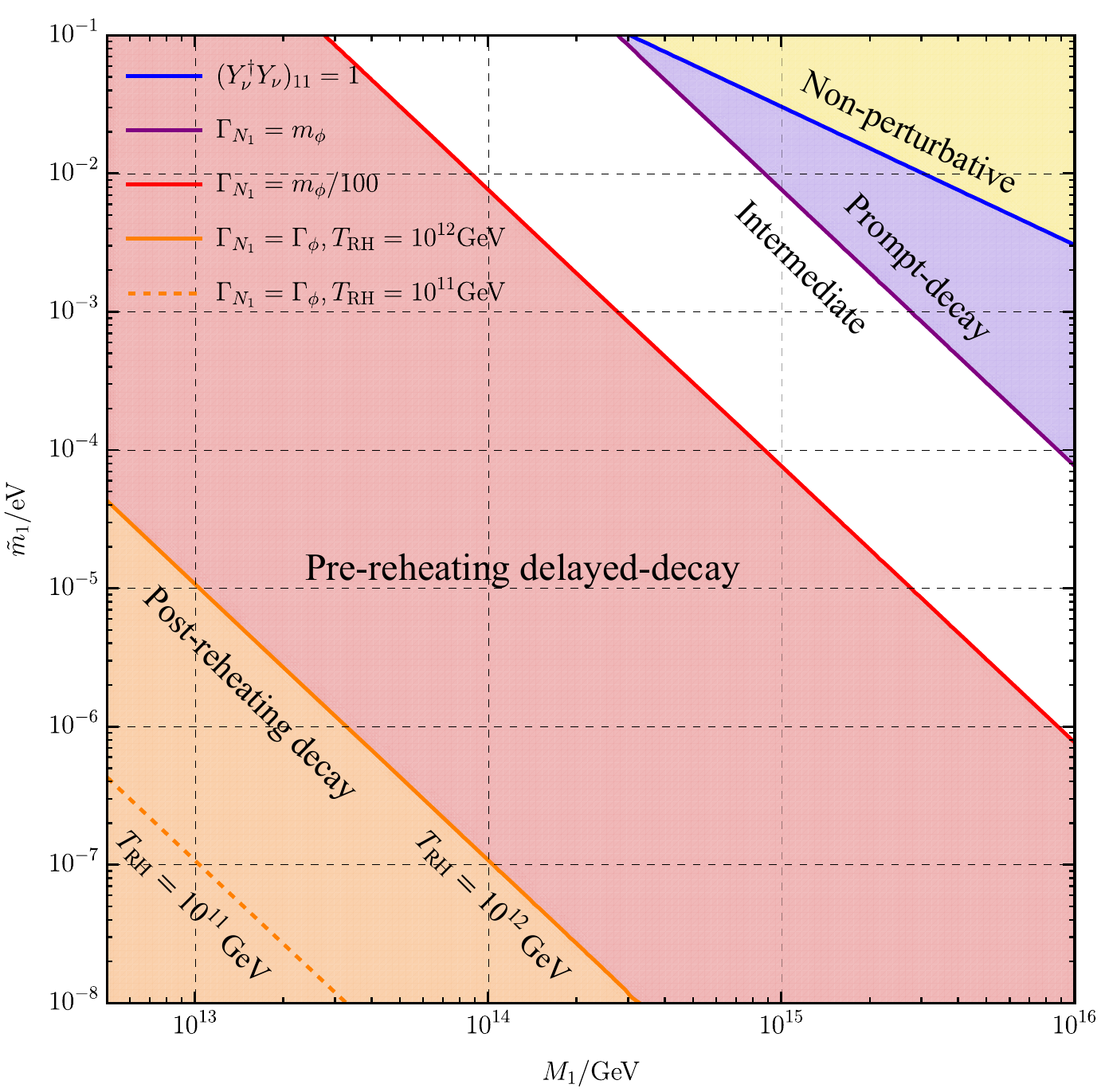}
\caption{
Parameter space of non-thermal leptogenesis for the benchmark values $T_{\rm RH}=10^{12}\,\mathrm{GeV}$ and $m_\phi=10^{13}\,\mathrm{GeV}$. 
The orange solid line denotes $\Gamma_{N_1}=\Gamma_\phi$, separating the region with $\Gamma_{N_1}<\Gamma_\phi$, where $N_1$ decays after reheating is completed, from the region with $\Gamma_{N_1}>\Gamma_\phi$, where $N_1$ decays before reheating occurs. 
The red solid line denotes $\Gamma_{N_1}=m_\phi/100$; below this line, $N_1$ survives over many inflaton oscillations before it decays. 
The purple solid line denotes $\Gamma_{N_1}=m_\phi$, above which the prompt-decay regime applies. 
The blue curve shows the perturbativity bound $(Y_\nu^\dagger Y_\nu)_{11}=1$. 
Four physically distinct regions are identified:
\textbf{prompt decay} ($\Gamma_{N_1}>m_\phi$, purple), where $N_1$ decays within one inflaton oscillation period;
\textbf{pre-reheating delayed decay} ($\Gamma_\phi<\Gamma_{N_1}\lesssim m_\phi/100$, red), where the non-thermal $N_1$ abundance builds up over many oscillations before $N_1$ decays prior to reheating;
\textbf{post-reheating decay} ($\Gamma_{N_1}<\Gamma_\phi$, orange), where $N_1$ survives until reheating is completed and subsequently decays in the radiation-dominated era.
All non-thermal regions must also satisfy the $\Delta L=2$ washout bound, which is automatically fulfilled for $T_{\rm RH}=10^{12}\,\mathrm{GeV}$ with realistic light-neutrino masses.
}
\label{fig:paramspace}
\end{figure}

The corresponding parameter space is displayed in Fig.~\ref{fig:paramspace} for the representative values $T_{\rm RH}=10^{12}\,\mathrm{GeV}$ and $m_\phi=10^{13}\,\mathrm{GeV}$. 
The $N_1$ decay rate is given by Eq.~\eqref{eq:GammaN1_mtilde}, while the inflaton decay rate is related to the reheating temperature through Eq.~\eqref{eq:Gamma_phi_TRH}. 
The conditions $\Gamma_{N_1}=m_\phi$, $\Gamma_{N_1}=m_\phi/100$, and $\Gamma_{N_1}=\Gamma_\phi$ then define simple contours in the $(M_1,\widetilde m_1)$ plane.

The colored regions in Fig.~\ref{fig:paramspace} show the three non-thermal decay regimes introduced above, together with the thermalized region excluded from the present analysis. 
The \textbf{prompt-decay} region (purple) satisfies $\Gamma_{N_1}>m_\phi$. 
In this regime, each preheating burst produces a nonzero $N_1$ abundance, but the produced particles decay within one inflaton oscillation period. 
Thus different bursts contribute independently to the final asymmetry, without forming a saturated $N_1$ population over many oscillations.

The \textbf{pre-reheating delayed-decay} region (red) satisfies $\Gamma_\phi<\Gamma_{N_1}\lesssim m_\phi/100$, together with the non-thermal condition $M_1>T_{\rm RH}$. 
In this regime, $N_1$ survives over many inflaton oscillations, so its abundance can build up and eventually saturate due to Pauli blocking. 
It then decays before the inflaton decay completes, namely before reheating occurs.

The \textbf{post-reheating decay} region (orange) corresponds to $\Gamma_{N_1}<\Gamma_\phi$ and $M_1>T_{\rm RH}$. 
Here the produced $N_1$ particles remain present until reheating is completed and subsequently decay in the radiation-dominated era.

The white region in Fig.~\ref{fig:paramspace} corresponds to an intermediate regime, where the lifetime of $N_1$ is neither short enough for prompt decay after each production burst nor long enough for the abundance to reach saturation before decay. In this regime, the final asymmetry depends sensitively on the detailed decay time of $N_1$. Its treatment is conceptually similar to the prompt-decay regime, but with a finite survival time after each production burst, so that the particles produced in one burst may partially affect later bursts. Since this requires a dedicated numerical evolution of both the occupation numbers and the decay process, we do not analyze this region in detail in the present work. Nevertheless, it can be incorporated straightforwardly in a numerical treatment.

All non-thermal regions must also satisfy the $\Delta L=2$ washout bound. 
This bound requires the $\Delta L=2$ scattering rate in the thermal bath after reheating to be smaller than the Hubble expansion rate, so that the lepton asymmetry generated from the non-thermally produced $N_1$ population is not erased. 
For the benchmark value $T_{\rm RH}=10^{12}\,\mathrm{GeV}$, this condition is automatically satisfied for realistic light-neutrino masses and therefore does not further reduce the allowed non-thermal regions in Fig.~\ref{fig:paramspace}. 
For a higher reheating temperature, the bound becomes stronger, especially for a quasi-degenerate light-neutrino spectrum.

In addition, the derivative interaction in Eq.~\eqref{eq:Lagrangian} should be interpreted as part of an effective field theory with cutoff $\Lambda\sim 2\pi f$. For consistency, the characteristic physical momentum of the produced fermions and the right-handed-neutrino mass should remain below this cutoff,
\begin{equation}
    \frac{k_{r,\max}}{a(t_{\rm max})}\lesssim \Lambda=2\pi f,\qquad M_1\lesssim \Lambda=2\pi f .
\end{equation}
where $a(t_{\rm max})\approx 1.5$ with $t_{\rm max} = 1.025\, m_{\phi}^{-1}$ being the time with the largest momentum ${k_{r,\max}}$ produced which is shown from Fig.~\ref{fig:ktilde} and Fig.~\ref{fig:adotphi}. Using the estimate \(k_{r,\max}\sim m_\phi M_{\rm Pl}/f\), this gives
\begin{equation}\label{eq:EFT_limit_k}
    m_\phi\frac{M_{\rm Pl}}{a(t_{\rm max}) f}\lesssim 2\pi f
\end{equation}
From the above equation, it directly follows that $f\gtrsim\sqrt{m_\phi M_{\rm Pl}/ (2 \pi a(t_{\rm max}))}\simeq 1.6 \times 10^{15}\,\mathrm{GeV}$. This sets a limit on $M_{\rm Pl}/f < 1500$. The other condition $M_1\lesssim \Lambda$ can be easily satisfied in the parameter space we consider.

\subsection{Pre-reheating delayed decay}\label{sec:case_preRH}

We first consider the pre-reheating delayed-decay regime. In this regime, the right-handed neutrinos survive over many inflaton oscillations, so that their non-thermal abundance can build up through repeated preheating production and eventually saturate due to Pauli blocking. At the same time, they decay before the inflaton has completed its decay, namely before reheating occurs. The relevant hierarchy of rates is therefore
\begin{equation}\label{eq:caseb_condition}
    \Gamma_\phi<\Gamma_{N_1}\ll m_\phi .
\end{equation}
In practice, to ensure that the saturated preheating abundance has already been reached before $N_1$ decays, we impose the more restrictive condition
\begin{equation}
    \Gamma_{N_1}\lesssim \frac{m_\phi}{100}.
\end{equation}
Using Eq.~\eqref{eq:GammaN1_mtilde}, this gives the upper bound
\begin{equation}\label{eq:mtilde_upper_caseb}
\widetilde m_1
<
\left(\frac{m_\phi}{10^{13}\,{\rm GeV}}\right)
\left(\frac{10^{14}\,{\rm GeV}}{M_1}\right)^2
\times 7.67\times 10^{-3}\,{\rm eV}.
\end{equation}
On the other hand, requiring $N_1$ to decay before reheating occurs gives
\begin{equation}
    \Gamma_{N_1}>\Gamma_\phi .
\end{equation}
Combining Eq.~\eqref{eq:GammaN1_mtilde} with Eq.~\eqref{eq:Gamma_phi_TRH}, we obtain the lower bound
\begin{equation}\label{eq:mtilde_lower_caseb}
\widetilde m_1
>
\left(\frac{T_{\rm RH}}{M_1}\right)^2
\times 1.08\times 10^{-3}\,{\rm eV}.
\end{equation}
In addition, to keep the right-handed-neutrino population genuinely non-thermal, we impose
\begin{equation}\label{eq:TRH_nonthermal_caseb}
    T_{\rm RH}<M_1 .
\end{equation}

We now relate the saturated right-handed-neutrino abundance to the final baryon asymmetry. The relevant input from the preheating calculation is the saturated comoving number density $N_{N_1}^{\rm pre}$. At the saturation time $t_{\rm sat}$, the corresponding physical number density is
\begin{equation}\label{eq:nN1_sat}
n_{N_1}^{\rm pre}(t_{\rm sat})
=
\frac{1}{a^3(t_{\rm sat})}
\int \frac{\text{d}^3k}{(2\pi)^3}
\left[
|\beta_{-,\rm pre}(k,t_{\rm sat})|^2
+
|\beta_{+,\rm pre}(k,t_{\rm sat})|^2
\right]
\equiv
\frac{N_{N_1}^{\rm pre}}{a^3(t_{\rm sat})}.
\end{equation}
Here $t_{\rm sat}$ denotes the time at which the occupation-number distribution has reached saturation and the comoving number density becomes approximately time independent. Numerically, as discussed in Sec.~\ref{sec.321}, this occurs after a characteristic time
\begin{equation}
    t_{\rm sat}\sim \mathcal{O}(10^2)\,m_\phi^{-1}.
\end{equation}
In the numerical estimates below, we evaluate the abundance at
\begin{equation}
    t_{\rm sat}=300\,m_\phi^{-1},
\end{equation}
which is sufficiently late for the occupation-number distribution and the corresponding comoving number density to have reached saturation. After saturation, $N_{N_1}^{\rm pre}$ remains approximately conserved until the decay of $N_1$, provided that the produced right-handed neutrinos remain subdominant in the total energy density.

Before reheating, the Universe is dominated by the coherent oscillation of the inflaton field and is effectively matter dominated. Hence,
\begin{equation}
    a^3\rho_\phi \simeq {\rm const.}
\end{equation}
Therefore, after the $N_1$ abundance has saturated, the ratio
\begin{equation}\label{eq:nrho_pre}
\left(\frac{n_{N_1}}{\rho_\phi}\right)_{\rm pre}
\equiv
\frac{n_{N_1}^{\rm pre}(t_{\rm sat})}{\rho_\phi(t_{\rm sat})}
=
\frac{N_{N_1}^{\rm pre}}
{a^3(t_{\rm sat})\rho_\phi(t_{\rm sat})}
\end{equation}
is conserved until the decay of $N_1$.

\begin{figure}[t]
    \centering
    \includegraphics[width=0.62\linewidth]{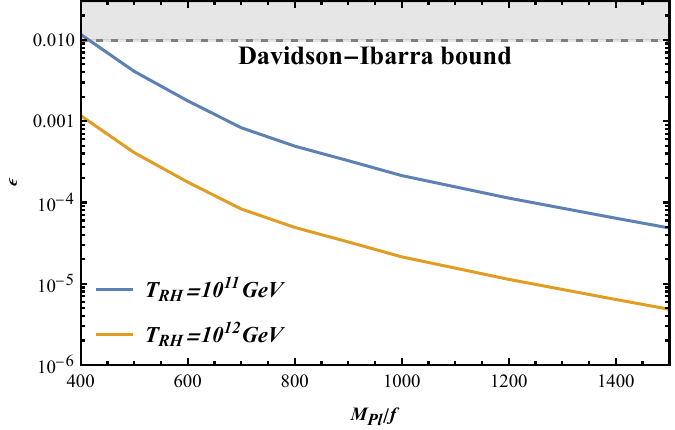}
    \caption{
    Required CP asymmetry $\epsilon$ for successful baryogenesis in the pre-reheating delayed-decay regime as a function of the coupling $M_{\rm Pl}/f$. 
    The benchmark parameters are $m_\phi=10^{13}\,\mathrm{GeV}$ and $M_1=10^{14}\,\mathrm{GeV}$.
    The horizontal dashed line shows the Davidson-Ibarra upper bound for the corresponding benchmark, which is given in~\ref{sec:DI}. The gray-shaded region is therefore forbidden by the Davidson-Ibarra upper bound. 
    Different curves correspond to different reheating temperatures, with the observed baryon asymmetry fixed to $n_B/s=8.79\times10^{-11}$. 
    A larger reheating temperature or a larger non-thermal $N_1$ abundance reduces the CP asymmetry required to reproduce the observed baryon asymmetry.
    }
    \label{fig:eps_caseb}
\end{figure}

Each $N_1$ decay generates a net lepton asymmetry $\epsilon$. Since in the present regime $N_1$ decays before reheating occurs, the generated lepton number is produced during the inflaton-oscillation era. Assuming that washout effects are negligible before reheating, this lepton number density simply redshifts as $a^{-3}$ until reheating. Equivalently, the ratio $n_L/\rho_\phi$ remains conserved before reheating. Thus,
\begin{equation}
    n_L = \epsilon\, n_{N_1},
\end{equation}
and
\begin{equation}\label{eq:nLrho_caseb}
\left(\frac{n_L}{\rho_\phi}\right)
=
\epsilon
\left(\frac{n_{N_1}}{\rho_\phi}\right)_{\rm pre}.
\end{equation}
When reheating occurs, the entropy density is
\begin{equation}
    s(T_{\rm RH})
    =
    \frac{2\pi^2}{45}g_\ast T_{\rm RH}^3,
\end{equation}
while the inflaton energy density is converted into radiation,
\begin{equation}
    \rho_\phi(T_{\rm RH})
    =
    \frac{\pi^2}{30}g_\ast T_{\rm RH}^4.
\end{equation}
Therefore,
\begin{equation}
    s(T_{\rm RH})
    =
    \frac{4}{3T_{\rm RH}}\rho_\phi(T_{\rm RH}).
\end{equation}
Using the conservation of $n_L/\rho_\phi$ between the $N_1$ decay and reheating, we obtain
\begin{equation}\label{eq:YL_caseb}
\frac{n_L}{s}
=
\frac{3}{4}\,
\epsilon\,T_{\rm RH}
\left(\frac{n_{N_1}}{\rho_\phi}\right)_{\rm pre}.
\end{equation}
The lepton asymmetry is partially converted into baryon number by electroweak sphaleron processes,
\begin{equation}
    B=\bar a(B-L),
    \qquad
    \bar a=\frac{28}{79}.
\end{equation}
This gives
\begin{equation}\label{eq:YB_caseb}
\frac{n_B}{s}
=
\bar a\,\frac{n_L}{s}
=
\frac{3}{4}\,
\bar a\,\epsilon\,T_{\rm RH}
\left(\frac{n_{N_1}}{\rho_\phi}\right)_{\rm pre}.
\end{equation}
Requiring agreement with the observed baryon asymmetry~\cite{Planck:2018vyg},
\begin{equation}
    \frac{n_B}{s}\simeq 8.79\times10^{-11},
\end{equation}
we find
\begin{equation}\label{eq:epsTRH_caseb}
\epsilon\,T_{\rm RH}
=
\frac{4}{3\bar a}
\left(\frac{\rho_\phi}{n_{N_1}}\right)_{\rm pre}
\times 8.79\times10^{-11}.
\end{equation}
As a representative benchmark, we take
\begin{equation}
    m_\phi=10^{13}\,{\rm GeV},
    \qquad
    M_1=10\,m_\phi=10^{14}\,{\rm GeV} .
\end{equation}
For each value of $M_{\rm Pl}/f$, the saturated comoving abundance $N_{N_1}^{\rm pre}$ is obtained from the numerical preheating calculation and inserted into Eq.~\eqref{eq:epsTRH_caseb}. The resulting value of the CP asymmetry required for successful baryogenesis is shown in Fig.~\ref{fig:eps_caseb} for several choices of the reheating temperature.

The behavior in Fig.~\ref{fig:eps_caseb} follows directly from Eq.~\eqref{eq:epsTRH_caseb}. For a fixed saturated right-handed-neutrino abundance, increasing $T_{\rm RH}$ enhances the final entropy-normalized baryon yield and therefore lowers the required value of $\epsilon$. This explains why the curves with larger reheating temperatures lie below those with smaller reheating temperatures. On the other hand, increasing $M_{\rm Pl}/f$ strengthens the axion--fermion coupling and enhances the non-thermal production of right-handed neutrinos during preheating. As a result, the ratio $(n_{N_1}/\rho_\phi)_{\rm pre}$ increases, and a smaller CP asymmetry is sufficient to reproduce the observed baryon asymmetry. This accounts for the downward trend of the required $\epsilon$ as $M_{\rm Pl}/f$ increases. We have also checked that, for all parameter points shown, the energy density stored in the produced $N_1$ population remains well below the inflaton energy density.

\subsection{Post-reheating decay}\label{sec:afterRH}

We now turn to the second delayed-decay regime, in which the non-thermally produced $N_1$ population survives until after reheating is completed. In this regime, the right-handed neutrinos decay more slowly than the inflaton,
\begin{equation}\label{eq:case_afterRH}
    \Gamma_{N_1}<\Gamma_\phi .
\end{equation}
The inflaton has therefore completed its decay when the $N_1$ population is still present, and the subsequent decay of $N_1$ takes place in the radiation-dominated era.

Combining Eq.~\eqref{eq:case_afterRH} with Eqs.~\eqref{eq:GammaN1_mtilde} and \eqref{eq:Gamma_phi_TRH}, we obtain the upper bound
\begin{equation}\label{eq:mtilde_upper_afterRH}
    \widetilde m_1
    <
    \left(\frac{T_{\rm RH}}{M_1}\right)^2
    \times 1.08\times 10^{-3}\,{\rm eV}.
\end{equation}
As in the previous regimes, we also impose the non-thermal condition
\begin{equation}\label{eq:TRH_nonthermal_afterRH}
    T_{\rm RH}<M_1 ,
\end{equation}
so that the right-handed neutrinos are not efficiently regenerated from the thermal bath after reheating. In addition, all non-thermal regions must satisfy the $\Delta L=2$ washout bound derived above. For the benchmark value $T_{\rm RH}=10^{12}\,{\rm GeV}$, this bound is automatically satisfied for realistic light-neutrino masses.

We now relate the saturated non-thermal abundance of $N_1$ to the final baryon asymmetry. The relevant input is the same saturated preheating abundance defined in Eq.~\eqref{eq:nN1_sat}. After saturation, the comoving number density $N_{N_1}^{\rm pre}$ remains approximately conserved until the decay of $N_1$, provided that the produced right-handed neutrinos remain subdominant in the total energy density. During the inflaton-oscillation era, the background is effectively matter dominated, so both $n_{N_1}$ and $\rho_\phi$ scale as $a^{-3}$ after saturation. Hence the ratio
\begin{equation}\label{eq:nrho_afterRH}
    \left(\frac{n_{N_1}}{\rho_\phi}\right)_{\rm pre}
\end{equation}
defined in Eq.~\eqref{eq:nrho_pre} is conserved from saturation to reheating.

At reheating, the inflaton energy density is converted into radiation,
\begin{equation}
    \rho_{\rm rad}(T_{\rm RH})
    =
    \frac{\pi^2}{30}g_\ast T_{\rm RH}^4,
    \qquad
    s(T_{\rm RH})
    =
    \frac{2\pi^2}{45}g_\ast T_{\rm RH}^3 .
\end{equation}
Using the conservation of $(n_{N_1}/\rho_\phi)_{\rm pre}$ before reheating, the $N_1$ number density at the completion of reheating is
\begin{equation}\label{eq:nN1_TRH_afterRH}
    n_{N_1}(T_{\rm RH})
    =
    \rho_{\rm rad}(T_{\rm RH})
    \left(\frac{n_{N_1}}{\rho_\phi}\right)_{\rm pre}.
\end{equation}
Therefore, the entropy-normalized $N_1$ abundance after reheating is
\begin{equation}\label{eq:YN1_afterRH}
    Y_{N_1}
    \equiv
    \frac{n_{N_1}}{s}
    =
    \frac{3}{4}\,
    T_{\rm RH}
    \left(\frac{n_{N_1}}{\rho_\phi}\right)_{\rm pre}.
\end{equation}
Since $N_1$ decays after reheating in the present regime, this yield remains conserved until $N_1$ decays, assuming no significant entropy production before its decay.

Each $N_1$ decay generates a net lepton asymmetry $\epsilon$. Neglecting washout effects, the lepton yield generated by the late decay is
\begin{equation}\label{eq:YL_afterRH}
    \frac{n_L}{s}
    =
    \epsilon\,Y_{N_1}
    =
    \frac{3}{4}\,
    \epsilon\,T_{\rm RH}
    \left(\frac{n_{N_1}}{\rho_\phi}\right)_{\rm pre}.
\end{equation}
The electroweak sphaleron processes partially convert this lepton asymmetry into baryon number,
\begin{equation}
    B=\bar a(B-L),
    \qquad
    \bar a=\frac{28}{79}.
\end{equation}
Thus the final baryon asymmetry is
\begin{equation}\label{eq:YB_afterRH}
    \frac{n_B}{s}
    =
    \bar a\,\frac{n_L}{s}
    =
    \frac{3}{4}\,
    \bar a\,\epsilon\,T_{\rm RH}
    \left(\frac{n_{N_1}}{\rho_\phi}\right)_{\rm pre}.
\end{equation}
This expression has the same form as in the pre-reheating delayed-decay regime, provided that the $N_1$ population remains subdominant and does not generate significant entropy upon decay.
The reason is that, in both regimes, the final baryon yield is controlled by the same saturated preheating abundance, encoded in $(n_{N_1}/\rho_\phi)_{\rm pre}$. The difference between the two regimes lies in the lifetime condition imposed on $\Gamma_{N_1}$.

Requiring agreement with the observed baryon asymmetry,
\begin{equation}
    \frac{n_B}{s}\simeq 8.79\times10^{-11},
\end{equation}
we obtain
\begin{equation}\label{eq:epsTRH_afterRH}
    \epsilon\,T_{\rm RH}
    =
    \frac{4}{3\bar a}
    \left(\frac{\rho_\phi}{n_{N_1}}\right)_{\rm pre}
    \times 8.79\times10^{-11}.
\end{equation}
Therefore, for the same preheating abundance, the required value of $\epsilon$ is identical to that in the pre-reheating delayed-decay regime. However, the allowed range of $\widetilde m_1$ is different, because the present regime requires $\Gamma_{N_1}<\Gamma_\phi$, or equivalently Eq.~\eqref{eq:mtilde_upper_afterRH}.

For the benchmark values used in Fig.~\ref{fig:paramspace}, namely $T_{\rm RH}=10^{12}\,{\rm GeV}$ and $m_\phi=10^{13}\,{\rm GeV}$, the post-reheating decay regime corresponds to the orange-shaded region with $M_1>T_{\rm RH}$ and $\Gamma_{N_1}<\Gamma_\phi$. In this region, the $N_1$ abundance is generated non-thermally during inflation and preheating, survives until reheating is completed, and then decays in the radiation-dominated era.

\paragraph{$N_1$ domination.}

In the post-reheating decay regime, there is an additional consistency condition. Since $N_1$ is non-relativistic after reheating, its energy density scales as $a^{-3}$, whereas the radiation energy density scales as $a^{-4}$. Therefore, even if the non-thermally produced $N_1$ population is initially subdominant at reheating, it may come to dominate the total energy density if it decays too late. Such a period of $N_1$ domination would modify the assumed radiation-dominated expansion history and could also lead to additional entropy production from $N_1$ decay. We therefore require that $N_1$ remains subdominant until it decays.

Let $T_D$ denote the temperature at which $N_1$ decays, defined by
\begin{equation}\label{eq:TD_def_afterRH}
    H(T_D)=\Gamma_{N_1}.
\end{equation}
During radiation domination,
\begin{equation}\label{eq:TD_afterRH}
    \Gamma_{N_1}
    =
    \sqrt{\frac{\pi^2g_\ast}{90}}\,
    \frac{T_D^2}{M_{\rm Pl}},
\end{equation}
so that $T_D<T_{\rm RH}$ follows from $\Gamma_{N_1}<\Gamma_\phi$.

After reheating, the $N_1$ yield is conserved until decay,
\begin{equation}
    Y_{N_1}
    =
    \frac{3}{4}\,
    T_{\rm RH}
    \left(\frac{n_{N_1}}{\rho_\phi}\right)_{\rm pre}.
\end{equation}
At $T=T_D$, the ratio of the $N_1$ energy density to the radiation energy density is
\begin{equation}\label{eq:N1dom_ratio_afterRH}
\frac{\rho_{N_1}(T_D)}{\rho_{\rm rad}(T_D)}
=
\frac{M_1Y_{N_1}s(T_D)}
{\rho_{\rm rad}(T_D)}
=
\frac{4}{3}\frac{M_1Y_{N_1}}{T_D}
=
\frac{M_1T_{\rm RH}}{T_D}
\left(\frac{n_{N_1}}{\rho_\phi}\right)_{\rm pre}.
\end{equation}
Requiring $N_1$ to remain subdominant before it decays gives
\begin{equation}\label{eq:N1dom_condition_afterRH}
    \frac{M_1T_{\rm RH}}{T_D}
    \left(\frac{n_{N_1}}{\rho_\phi}\right)_{\rm pre}
    <1 .
\end{equation}
Equivalently,
\begin{equation}\label{eq:TD_lower_afterRH}
    T_D
    >
    M_1T_{\rm RH}
    \left(\frac{n_{N_1}}{\rho_\phi}\right)_{\rm pre}.
\end{equation}
Using Eq.~\eqref{eq:TD_afterRH}, this condition can be rewritten as a lower bound on the decay rate,
\begin{equation}\label{eq:GammaN_dom_afterRH}
    \Gamma_{N_1}
    >
    \sqrt{\frac{\pi^2g_\ast}{90}}\,
    \frac{1}{M_{\rm Pl}}
    \left[
    M_1T_{\rm RH}
    \left(\frac{n_{N_1}}{\rho_\phi}\right)_{\rm pre}
    \right]^2 .
\end{equation}
In terms of the effective neutrino mass parameter, using Eq.~\eqref{eq:GammaN1_mtilde}, this becomes
\begin{equation}\label{eq:mtilde_dom_afterRH}
    \widetilde m_1
    >
    \frac{4\pi v^2}{M_1^2}
    \sqrt{\frac{\pi^2g_\ast}{90}}\,
    \frac{1}{M_{\rm Pl}}
    \left[
    M_1T_{\rm RH}
    \left(\frac{n_{N_1}}{\rho_\phi}\right)_{\rm pre}
    \right]^2 .
\end{equation}
For the parameter space considered here, we have checked numerically that the non-thermal abundance is typically small enough that this condition is easily satisfied. Nevertheless, it provides an additional consistency check for the post-reheating decay regime. We have also checked that the total energy density stored in the produced $N_1$ population remains well below the inflaton energy density for all parameter points shown.

\begin{figure}[t]
    \centering
    \includegraphics[width=0.62\linewidth]{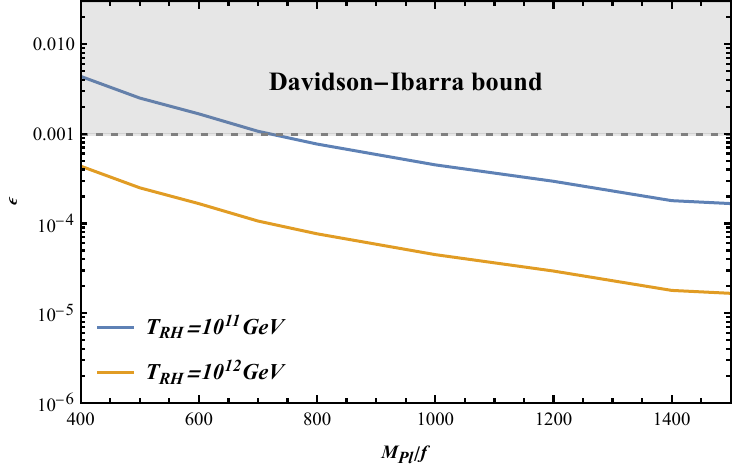}
    \caption{
    Required CP asymmetry $\epsilon$ for successful baryogenesis in the post-reheating decay regime as a function of the coupling $M_{\rm Pl}/f$. 
    The benchmark parameters are $m_\phi=10^{13}\,\mathrm{GeV}$ and $M_1=10^{13}\,\mathrm{GeV}$. 
    The horizontal dashed line shows the Davidson-Ibarra upper bound for the corresponding benchmark, which is given in~\ref{sec:DI}. The gray-shaded region is therefore forbidden by the Davidson-Ibarra upper bound. 
    Different curves correspond to different reheating temperatures, with the observed baryon asymmetry fixed to $n_B/s=8.79\times10^{-11}$. 
    A larger reheating temperature or a larger non-thermal $N_1$ abundance reduces the CP asymmetry required to reproduce the observed baryon asymmetry.
    }
    \label{fig:eps_casea}
\end{figure}

As a representative benchmark, we take
\begin{equation}
    m_\phi=10^{13}\,\mathrm{GeV},\quad M_1=10^{13}\,\mathrm{GeV} .
\end{equation}
Since the baryon-yield formula has the same form as in the pre-reheating delayed-decay regime, the difference here arises from the lifetime condition $\Gamma_{N_1}<\Gamma_\phi$ and from the benchmark choice of $M_1$. The resulting CP asymmetry required for successful baryogenesis is shown in Fig.~\ref{fig:eps_casea}.

\subsection{Prompt decay}\label{sec:case_c}

We finally consider the prompt-decay regime, corresponding to regime~\ref{case_c}. In this regime, the right-handed neutrinos decay within one inflaton oscillation period,
\begin{equation}\label{eq:casec_condition}
    \Gamma_{N_1}>m_\phi .
\end{equation}
The particles produced in a given non-adiabatic burst therefore decay before the next burst becomes relevant. As a result, different production bursts contribute approximately independently to the final lepton asymmetry. This differs from the delayed-decay regimes discussed above, where the occupation-number distribution can build up over many oscillations and eventually saturate due to Pauli blocking.

Using Eq.~\eqref{eq:GammaN1_mtilde}, the prompt-decay condition gives the lower bound
\begin{equation}\label{eq:mtilde_lower_casec}
\widetilde m_1
>
\left(\frac{m_\phi}{10^{13}\,{\rm GeV}}\right)
\left(\frac{10^{14}\,{\rm GeV}}{M_1}\right)^2
\times 0.767\,{\rm eV}.
\end{equation}
On the other hand, perturbativity of the neutrino Yukawa coupling requires
\begin{equation}
    (Y_\nu^\dagger Y_\nu)_{11}<1 .
\end{equation}
Using Eq.~\eqref{eq:mtilde}, this gives the upper bound
\begin{equation}\label{eq:mtilde_upper_casec}
\widetilde m_1
<
\left(\frac{10^{14}\,{\rm GeV}}{M_1}\right)
\times 0.305\,{\rm eV}.
\end{equation}
The corresponding parameter region is shown by the purple-shaded area in Fig.~\ref{fig:paramspace}. This region is viable only where the lower bound in Eq.~\eqref{eq:mtilde_lower_casec} is compatible with the perturbativity bound in Eq.~\eqref{eq:mtilde_upper_casec}. The condition $\Gamma_{N_1}>m_\phi$ should be understood as a conservative limiting regime that makes the stage-by-stage treatment reliable. A more general intermediate lifetime would require a coupled numerical evolution of production and decay, as discussed above.

In the prompt-decay regime, the inflationary contribution to the final asymmetry is strongly suppressed. Since the typical Hubble scale during inflation satisfies $H_{\rm inf}\simeq 0.5m_\phi$, the condition $\Gamma_{N_1}>m_\phi$ implies that the right-handed neutrinos produced during inflation decay within a time shorter than one Hubble time. Their contribution is then diluted by the subsequent expansion before reheating. We therefore focus on the right-handed neutrinos produced during preheating.

Let $\Delta N_{i,{\rm pre}}$ denote the comoving number density of right-handed neutrinos produced during the $i$th non-adiabatic production burst, evaluated after that burst. Since the produced particles decay promptly, each burst generates its own lepton asymmetry independently. The total comoving lepton number generated during preheating is therefore
\begin{equation}\label{eq:NL_casec_sum}
    N_L^{\rm pre}
    =
    \epsilon
    \sum_{i=1}^{n_{\rm count}}
    \Delta N_{i,{\rm pre}},
\end{equation}
where $n_{\rm count}$ denotes the number of production bursts that satisfy the non-adiabaticity condition. In practice, this number is determined by
\begin{equation}
    |\dot\theta|\gtrsim M_1 ,
\end{equation}
as discussed in Sec.~\ref{sec.322}.

Equivalently, the quantity entering the final baryon yield can be written as
\begin{equation}\label{eq:nrho_casec}
\left(\frac{n_{N_1}}{\rho_\phi}\right)_{\rm pre}^{\rm prompt}
\equiv
\sum_{i=1}^{n_{\rm count}}
\frac{\Delta N_{i,{\rm pre}}}
{a^3(t_i)\rho_\phi(t_i)} ,
\end{equation}
where $t_i$ denotes the time immediately after the $i$th production burst. During the inflaton-oscillation era, the background is effectively matter dominated, so $a^3\rho_\phi\simeq{\rm const.}$. Therefore, Eq.~\eqref{eq:nrho_casec} is equivalent to summing the comoving lepton asymmetry generated by all relevant bursts, evolving it to reheating, and normalizing it by the inflaton energy density at reheating.

Using
\begin{equation}
    s(T_{\rm RH})
    =
    \frac{4}{3T_{\rm RH}}\rho_\phi(T_{\rm RH}),
\end{equation}
the lepton asymmetry yield in the prompt-decay regime is
\begin{equation}\label{eq:YL_casec}
\frac{n_L}{s}
=
\frac{3}{4}\,
\epsilon\,T_{\rm RH}
\left(\frac{n_{N_1}}{\rho_\phi}\right)_{\rm pre}^{\rm prompt}.
\end{equation}
After sphaleron conversion, the baryon asymmetry becomes
\begin{equation}\label{eq:YB_casec}
\frac{n_B}{s}
=
\frac{3}{4}\,
\bar a\,\epsilon\,T_{\rm RH}
\left(\frac{n_{N_1}}{\rho_\phi}\right)_{\rm pre}^{\rm prompt},
\end{equation}
where $\bar a=28/79$.

Requiring agreement with the observed baryon asymmetry,
\begin{equation}
    \frac{n_B}{s}\simeq 8.79\times10^{-11},
\end{equation}
we obtain
\begin{equation}\label{eq:epsTRH_casec}
\epsilon\,T_{\rm RH}
=
\frac{4}{3\bar a}
\left[
\left(\frac{n_{N_1}}{\rho_\phi}\right)_{\rm pre}^{\rm prompt}
\right]^{-1}
\times 8.79\times10^{-11}.
\end{equation}

The number of relevant production bursts can be estimated from the condition that the oscillating background drives the helicity-dependent effective frequency through the non-adiabatic region. Up to order-one corrections from the detailed background evolution, we find approximately
\begin{equation}\label{eq:ncount_casec}
\begin{aligned}
n_{\rm count}
&\sim
\frac{M_{\rm Pl}m_\phi}{M_1 f},
\qquad
m_\phi \le M_1 \le \frac{M_{\rm Pl}m_\phi}{4f},
\\
n_{\rm count}
&=1,
\qquad
\frac{M_{\rm Pl}m_\phi}{4f}<M_1\le \frac{M_{\rm Pl}m_\phi}{f}.
\end{aligned}
\end{equation}
If the condition $|\dot\theta|\gtrsim M_1$ is never satisfied, no complete production burst occurs during preheating and $n_{\rm count}=0$, although partial non-adiabatic excitation may still be present. In practice, only the first few bursts contribute significantly, because the amplitude of the inflaton oscillation decreases with time.

Compared with the delayed-decay regimes, the prompt-decay regime can lead to a larger effective right-handed-neutrino abundance. This does not mean that the Pauli exclusion principle is violated. Rather, the occupation produced in one burst does not persist until the next burst: the produced right-handed neutrinos decay away before later non-adiabatic production occurs. Consequently, the suppression associated with previously occupied final states is strongly reduced, and similar momentum modes can be populated repeatedly in successive bursts. This can increase the total stage-summed abundance and reduce the CP asymmetry required for successful baryogenesis.

\begin{figure}[t]
    \centering
    \includegraphics[width=0.62\linewidth]{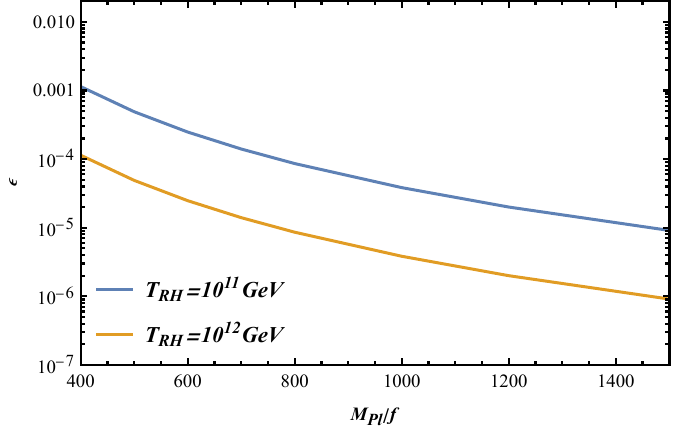}
    \caption{
    Required CP asymmetry $\epsilon$ for successful baryogenesis in the prompt-decay regime as a function of the coupling $M_{\rm Pl}/f$. 
    The benchmark mass is $M_1=5\times 10^{14}\,\mathrm{GeV}$. 
    Here the Davidson-Ibarra bound is not shown because it lies above the plotted range. 
    Different curves correspond to representative reheating temperatures, with the observed baryon asymmetry fixed to $n_B/s=8.79\times10^{-11}$. 
    In this regime, the relevant right-handed-neutrino abundance is obtained by summing the stage-by-stage preheating yields.
    }
    \label{fig:eps_casec}
\end{figure}

As a representative benchmark, we take
\begin{equation}
    M_1=50m_\phi=5\times 10^{14} \rm GeV~.
\end{equation}
For each value of $M_{\rm Pl}/f$, we compute the stage-summed abundance in Eq.~\eqref{eq:nrho_casec} and insert it into Eq.~\eqref{eq:epsTRH_casec}. The resulting CP asymmetry required for successful baryogenesis is shown in Fig.~\ref{fig:eps_casec} for several choices of the reheating temperature.

The behavior in Fig.~\ref{fig:eps_casec} follows from Eq.~\eqref{eq:epsTRH_casec}. A larger reheating temperature increases the final entropy-normalized baryon yield and therefore reduces the required CP asymmetry. Increasing $M_{\rm Pl}/f$ strengthens the axion--fermion coupling, making non-adiabatic fermion production more efficient and increasing the stage-summed ratio $\left(n_{N_1}/\rho_\phi\right)_{\rm pre}^{\rm prompt}$. As a result, a smaller value of $\epsilon$ is sufficient to reproduce the observed baryon asymmetry. For larger fermion masses, production is less efficient, and a stronger coupling is needed to obtain the same baryon asymmetry.

In the prompt decay regime, we also checked that the total energy stored in the produced $N_1$ population remains well below the inflaton energy density for all parameter points shown.

\paragraph{Davidson-Ibarra bound on the CP asymmetry}\label{sec:DI}

For hierarchical right-handed neutrinos, the CP asymmetry parameter $\epsilon$ of the lightest heavy neutrino $N_1$ is bounded from above by the Davidson-Ibarra bound~\cite{Davidson:2002qv,Buchmuller:2003gz}:
\[
|\epsilon| \le \epsilon_{\mathrm{DI}} \equiv \frac{3}{8\pi}\, \frac{M_1\, m_{\nu}^{\max}}{v^2},
\]
where $M_1$ is the mass of $N_1$, $v\approx 246\;\mathrm{GeV}$ is the Higgs vacuum expectation value, and $m_{\nu}^{\max}$ denotes the largest light neutrino mass, typically identified with the atmospheric neutrino mass scale $m_{\nu}^{\max}\simeq 0.05\;\mathrm{eV}$. Numerically,
\[
\epsilon_{\mathrm{DI}} \simeq 9.86\times 10^{-4}\left(\frac{M_1}{10^{13}\,\mathrm{GeV}}\right).
\]
This bound originates from the seesaw relation between the light neutrino masses and the Yukawa couplings, combined with the requirement that the CP asymmetry does not exceed the maximal value allowed by the neutrino oscillation data. In any viable leptogenesis scenario, the actual CP asymmetry $\epsilon$ must satisfy $|\epsilon|\le \epsilon_{\mathrm{DI}}$.

In summary, all three non-thermal regimes can in principle support successful leptogenesis driven by axion-induced fermion production. In the two delayed-decay regimes, the final baryon asymmetry is controlled by the saturated preheating abundance. The difference between them lies in whether $N_1$ decays before or after reheating is completed, which leads to different lifetime constraints on $\widetilde m_1$. In the prompt-decay regime, by contrast, the final asymmetry is controlled by the sum of the stage-by-stage production yields. Since the particles produced in one burst decay before the next burst becomes relevant, Pauli blocking from earlier bursts is reduced, and the same momentum range can contribute repeatedly. These results show that non-thermal production of heavy right-handed neutrinos can provide a viable origin of the baryon asymmetry in appropriate regions of parameter space through their CP-violating decays.

\section{Conclusion}
\label{sec:conclusion}

In this work, we have studied fermion production from an axion-like inflaton with a derivative coupling, with an emphasis on organizing the dynamics according to the full production history and clarifying the underlying physical mechanisms. Our approach provides a unified description of particle production during inflation and preheating, and allows for a transparent interpretation of how the resulting fermion abundance is built up over time.

During slow-roll inflation, the rolling inflaton background induces a helicity-asymmetric fermion spectrum. This asymmetry is controlled by the ratio between the effective coupling scale and the fermion mass, and leads to a characteristic momentum distribution that already provides a non-trivial initial condition for the subsequent evolution.

During preheating, the oscillating inflaton background drives repeated violations of adiabaticity. We have shown that these non-adiabatic events are governed by a helicity-dependent effective momentum, whose zero crossings provide a simple and physically transparent criterion for particle production. This picture naturally explains the appearance of a momentum cutoff in the fermion spectrum, as well as its helicity dependence.

A central result of this work is that the fermion production history depends qualitatively on the lifetime of the produced particles. In the delayed-decay regime, the occupation number accumulates over multiple production events and eventually saturates due to Pauli blocking, leading to a quasi-stationary non-thermal distribution. In contrast, in the prompt-decay regime, the production must be described in an event-based manner, where each non-adiabatic burst contributes independently and the total abundance is obtained by summing over successive stages. This distinction is essential for a consistent interpretation of the final fermion abundance.

We have applied these results to the case in which the produced fermion is identified with a heavy right-handed neutrino, and investigated the resulting non-thermal leptogenesis. In the delayed-decay regimes, the relevant input is the saturated abundance generated during preheating, while in the prompt-decay regime the baryon asymmetry is sourced by a sequence of production-and-decay events. We find that both regimes can reproduce the observed baryon asymmetry in viable regions of parameter space, but they differ parametrically: in particular, the prompt-decay regime typically requires a smaller CP asymmetry due to the absence of strong Pauli suppression across successive production stages.

More broadly, our analysis highlights that non-equilibrium particle production in time-dependent backgrounds can exhibit qualitatively different behaviors which depend sensitively on the interplay among production, decay, and quantum statistics. The framework developed here provides a systematic way to track these effects across different cosmological epochs and may be useful for studying other scenarios involving fermion production in the early Universe.

Several directions deserve further investigation. It would be interesting to incorporate a more complete treatment of the decay and thermalization processes, for example by embedding the production mechanism into a Boltzmann or density-matrix framework. In addition, extending the analysis to include interactions with other sectors, or exploring possible observational signatures associated with the helicity asymmetry, could further elucidate the phenomenological implications of this scenario.

\acknowledgments

This work is supported by the National Key R{\&}D Program of China under grant 2023YFA1606100 and by the National Natural Science Foundation of China under grants No. 12435005. C.\,H.\ acknowledges support from the Sun Yat-Sen University Science Foundation, 
the Fundamental Research Funds for the Central Universities at Sun Yat-sen University under Grant No.\,24qnpy117, and the Key Laboratory of Particle Astrophysics and Cosmology (MOE) of Shanghai Jiao Tong University.

\appendix
\section{Fermion Basis and the Definition of Occupation Number}\label{app1}
It is useful to clarify why we choose to perform the calculation of fermion occupation numbers in the $\psi$ basis rather than in the original $Y$ basis. Although the two descriptions are related through the chiral rotation $\psi\equiv e^{-i\gamma^5\theta}Y$ and therefore describe the same physical theory, they lead to different Hamiltonian formulations and, consequently, to different definitions of the fermion occupation number. As we will show below, in the small-mass regime the occupation number defined in the $Y$ basis is not well behaved and can give rise to unphysical results. By contrast, the occupation number defined in the Hamiltonian-diagonal $\psi$ basis remains physically well behaved throughout the evolution. In this appendix, we present a detailed analysis of this issue and explain why the $\psi$ basis provides the appropriate framework for our calculation.

Before discussing the Hamiltonian formulation and fermion occupation numbers, let us first compare the two field bases at the level of the Lagrangian. In the original $Y$ basis, the axion-fermion coupling appears as a derivative interaction with the axial fermion current,
\begin{equation}
\mathcal{L}_Y=
\bar{Y}
\left(
i\gamma^\mu\partial_\mu
-ma
+\gamma^\mu\gamma^5\partial_\mu\theta
\right)
Y,
\end{equation}
This form makes the derivative nature of the axion coupling manifest. However, it also makes the massless limit somewhat obscure: even when $m\to0$, the interaction term $\partial_\mu\theta\, \bar Y\gamma^\mu\gamma^5Y$ seems to remain present.

The same theory can be rewritten by an axial field redefinition. In the rotated $\psi$ basis, the derivative coupling is removed from the fermion kinetic term and is transferred into a time-dependent phase of the mass term,
\begin{equation}
\mathcal L_\psi=\bar\psi\left(i\gamma^\mu\partial_\mu-mae^{2i\gamma^5\theta}\right)\psi.
\end{equation}
In this form, all dependence on the axion field is proportional to the fermion mass. Therefore, in the limit $m\to0$, the coupling between the axion background and the fermion vanishes manifestly.

The two bases are of course physically equivalent. In the $Y$ basis, the disappearance of the axion-fermion coupling in the massless limit can also be made explicit by integrating the derivative interaction by parts and using the equation of motion. Up to a total derivative, the derivative coupling $\partial_\mu\theta\,\bar Y\gamma^\mu\gamma^5Y$ can be rewritten as
\begin{equation}
\partial_\mu\theta\,\bar Y\gamma^\mu\gamma^5Y
\simeq
-\theta\,\partial_\mu\left(\bar Y\gamma^\mu\gamma^5Y\right),
\end{equation}
where $\simeq$ denotes equality modulo a total derivative. Using the equation of motion~\eqref{eq:Y_eom}, one obtains
\begin{equation}
\partial_\mu\left(\bar Y\gamma^\mu\gamma^5Y\right)
=
2ima\,\bar Y\gamma^5Y ,
\end{equation}
up to possible anomaly terms. Therefore, the derivative interaction is equivalently given by
\begin{equation}
\partial_\mu\theta\,\bar Y\gamma^\mu\gamma^5Y
\simeq
-2ima\theta\,\bar Y\gamma^5Y .
\end{equation}
This form makes it explicit that the axion-fermion coupling is proportional to the fermion mass and hence vanishes in the massless limit. Nevertheless, this cancellation is not manifest at the level of the original Lagrangian. For the purpose of studying fermion production, especially in the small-mass regime, the $\psi$ basis is therefore more transparent: it makes the decoupling of the inflaton from a massless fermion explicit.

Next, we derive the Hamiltonian in the $Y$ basis and obtain the corresponding expression for the fermion occupation number. Using the definition of the Hamiltonian together with the equation of motion for the fermion field $Y$, one obtains the fermion Hamiltonian in the $Y$ basis
\begin{equation}
    H = i\int \mathrm{d}^3x\, Y^\dagger Y',
\end{equation}
Substituting the plane-wave expansion of $Y$ in Eq.~\eqref{eq:Y_fourier} and the mode-function equations of motion in Eq.~\eqref{eq:u_eq_Y}~\eqref{eq:v_eq_Y} into the above expression, and using the charge-conjugation relations in Eq.~\eqref{ccrelation} together with the helicity-spinor identities in Eq.~\eqref{helicity-spinor identities}, the Hamiltonian can be written as:
\begin{equation}\label{eq:psi_APP}
H
=
\int \mathrm{d}^3k
\sum_{r=\pm}
\begin{pmatrix}
\hat a_r^\dagger(\mathbf{k}) &
\hat a_r(-\mathbf{k})
\end{pmatrix}
\begin{pmatrix}
\tilde{A}_r(k,t) & \tilde{B}_r^*(k,t) \\
\tilde{B}_r(k,t) & -\tilde{A}_r(k,t)
\end{pmatrix}
\begin{pmatrix}
\hat a_r(\mathbf{k}) \\
\hat a_r^\dagger(-\mathbf{k})
\end{pmatrix},
\end{equation}
where
\begin{equation}\label{eq:Y_Hamiltonian_A_B}
\begin{aligned}
\tilde{A}_r&=
	\frac{1}{2}
	\left[
	-(kr-a\dot\theta)
	\left(
	|\tilde u_r|^2-|\tilde v_r|^2
	\right)
	+mar
	\left(
	\tilde u_r^*\tilde v_r+\tilde v_r^*\tilde u_r
	\right)
	\right]\\
\tilde{B}_r &= -r\frac{e^{ir\varphi_{\mathbf k}}}{2}
	\left[
	2(kr-a\dot\theta)\tilde u_r\tilde v_r
	+mar\left(\tilde u_r^2-\tilde v_r^2\right)
	\right].
\end{aligned}
\end{equation}
After performing the Bogolyubov transformation in Eq.~\eqref{eq:BT}, one obtains the Bogolyubov coefficients
\begin{equation}\label{Y_BC}
\begin{aligned}
\tilde{\alpha}_r
&=
\frac{e^{ir\varphi_{\mathbf k}/2}}{2}
\left[
\sqrt{1-\frac{kr-a\dot\theta}{\tilde{\omega}^r_k}}\,\tilde u_r
+
r\sqrt{1+\frac{kr-a\dot\theta}{\tilde{\omega}^r_k}}\,\tilde v_r
\right],
\\
\tilde{\beta}_r
&=
\frac{e^{ir\varphi_{\mathbf k}/2}}{2}
\left[
\sqrt{1+\frac{kr-a\dot\theta}{\tilde{\omega}^r_k}}\,\tilde u_r
-
r\sqrt{1-\frac{kr-a\dot\theta}{\tilde{\omega}^r_k}}\,\tilde v_r
\right],
\end{aligned}
\end{equation}
and then the Hamiltonian can be written in diagonal form
\begin{equation}\label{eq:diag_H_APP}
H
=
\int \mathrm{d}^3k
\sum_{r=\pm}
\tilde{\omega}^r_k(t)
\left[
\hat A_r^\dagger(\mathbf{k},t)\hat A_r(\mathbf{k},t)
-
\hat A_r(-\mathbf{k},t)\hat A_r^\dagger(-\mathbf{k},t)
\right],
\end{equation}
where
\begin{equation}
    \tilde{\omega}^r_k(t) = \sqrt{(kr-a\dot{\theta})^2 +m^2a^2}
\end{equation}
is the instantaneous eigen-energy. It should be emphasized that, throughout this discussion, we neglect inflaton perturbations and approximate the inflaton only as a homogeneous, time-dependent background.

To simplify the notation, we define effective momentum
\begin{equation}
\tilde{k}_r(t)\equiv a(t)(p(t)r-\dot{\theta}) = a(t)(p(t)r+\xi H_{\rm inf}),
\end{equation}
where $p(t)\equiv k/a(t)$ denotes the physical momentum and is redshifted by the cosmic expansion, $\xi$ be defined in Eq.~\eqref{eq:mu_xi_x}. The eigenenergy can then be written as:
\begin{equation}
    \tilde{\omega}^r_k(t) = \sqrt{\tilde{k}_r^2(t)+m^2a^2(t)}.
\end{equation}
During inflation, $\xi>0$ can be approximated as a constant. For the negative-helicity mode, as $p(t)$ is redshifted by the cosmic expansion, the effective momentum can approach zero at a finite time. This can lead to unphysical results in the computation of the fermion occupation number, especially in the small-mass limit.

To make this point more transparent, we compute in the $Y$ basis the evolution of the fermion occupation number as the physical momentum is redshifted. This is done by substituting the inflationary solutions of the mode functions in Eq.~\eqref{u_solve} and~\eqref{v_solve} into Eq.~\eqref{Y_BC} and evaluating the occupation number $|\tilde{\beta}_r|^2$. We focus on the small-mass regime. For the positive-helicity fermion, the occupation number is always suppressed by a factor of order $\mathcal{O}(\mu^2)$, since $\tilde \beta_+\propto \tilde{u}_+\propto \mu\equiv m/H_{\rm inf}$ in the limit $\mu\to0$. This behavior is consistent with the physical expectation that the axion-fermion coupling vanishes in the massless limit, and hence fermion production should be suppressed.

\begin{figure}[htbp]
\centering
\includegraphics[width=.44\textwidth]{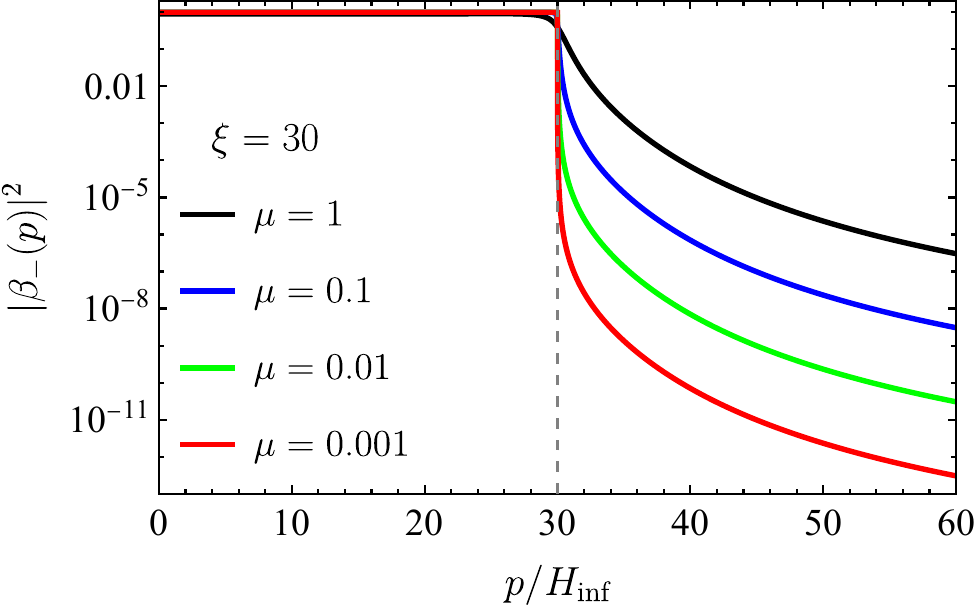}
\includegraphics[width=.44\textwidth]{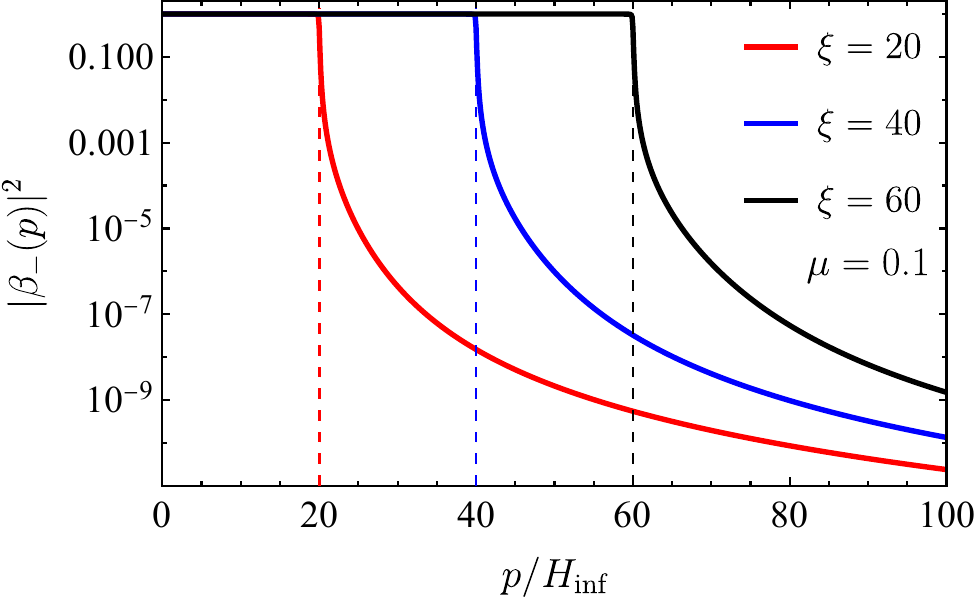}
\includegraphics[width=.44\textwidth]{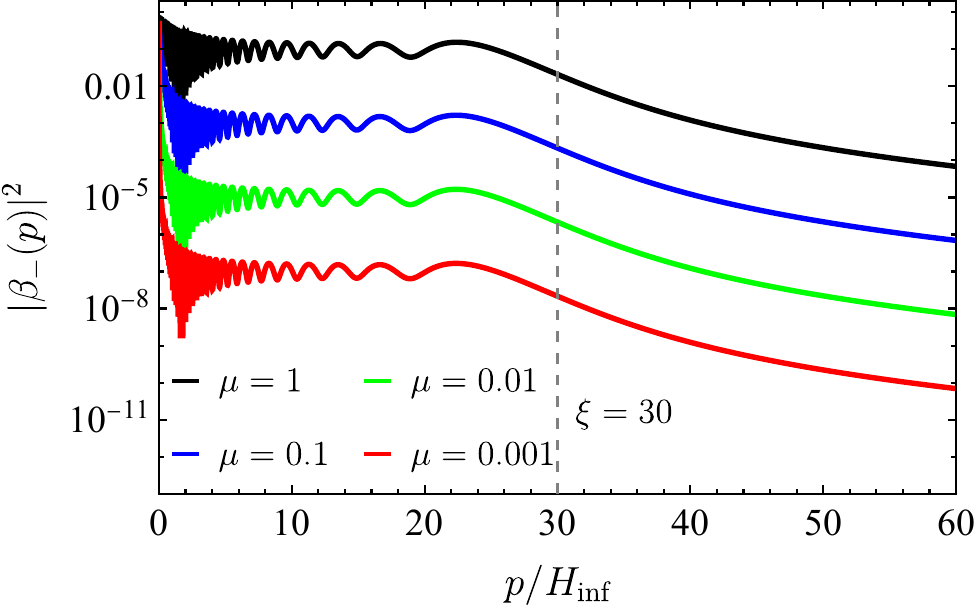}
\includegraphics[width=.44\textwidth]{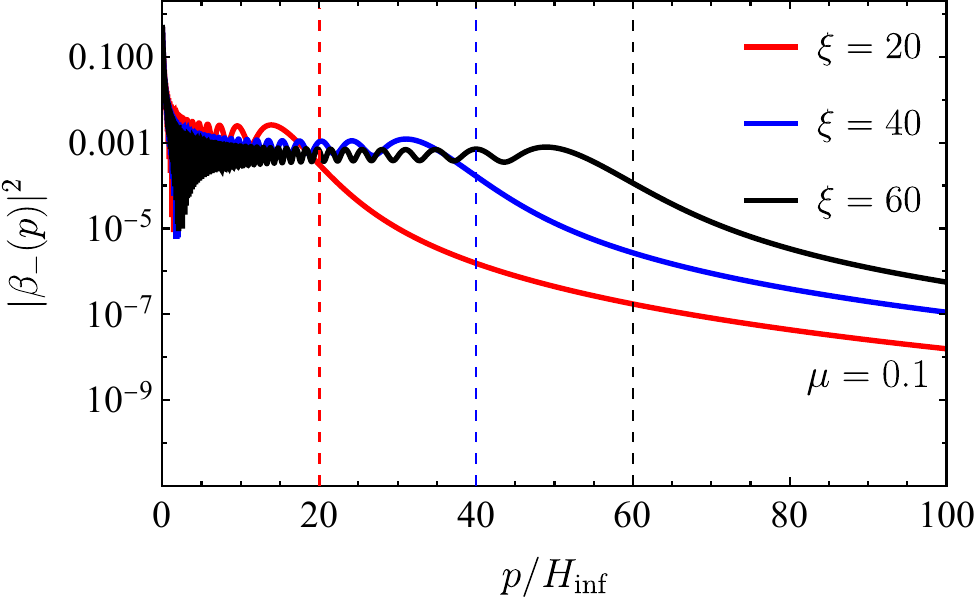}
\caption{
Evolution of the negative-helicity fermion occupation number $|\beta_-(p)|^2$ as a function of the physical momentum $p(t)=k/a(t)$ for a given mode $k$. 
The upper panels show the result defined in the $Y$ basis, while the lower panels show the corresponding result defined in the $\psi$ basis. 
The left column shows the dependence on the mass parameter $\mu\leq 1$ for fixed $\xi=30$, while the right column shows the dependence on the coupling parameter $\xi$ for fixed $\mu=0.1$. 
The vertical dashed lines indicate the point $p/H_{\rm inf}=\xi$, where the effective momentum of the negative-helicity mode $\tilde k_-(t)$ crosses zero in the $Y$ basis; the same lines are shown in the lower panels for comparison. 
}
\label{fig:appendix}
\end{figure}
For the negative-helicity fermion, the behavior is markedly different. In Fig.~\ref{fig:appendix}, we compare the evolution of the negative-helicity occupation number in the $Y$ and $\psi$ bases as a function of the physical momentum $p(t)$ for different values of the mass parameter $\mu\leq1$ and the coupling parameter $\xi$. The upper panels show the occupation number defined in the $Y$ basis, while the lower panels show the corresponding result in the $\psi$ basis. In the $Y$ basis, at early times when $p/H_{\rm inf}>\xi$, the occupation number is suppressed in the small-mass regime. This follows from $\tilde k_-/\tilde\omega_k^-\simeq -1$, for which the occupation number is dominated by $|\tilde v_-|^2\sim\mathcal{O}(\mu^2)$. However, as the physical momentum is redshifted and crosses $p/H_{\rm inf}\simeq\xi$, the ratio $\tilde k_-/\tilde\omega_k^-$ changes rapidly from $-1$ to $+1$. The instantaneous particle-number definition in the $Y$ basis is then dominated by $|u_-|^2\sim\mathcal{O}(1)$, leading to an abrupt jump of the occupation number from a mass-suppressed value to a value close to unity. For $p/H_{\rm inf}<\xi$, the $Y$-basis occupation number remains saturated near unity, almost independently of how small $\mu$ is.

This behavior is not reproduced in the $\psi$ basis. As shown in the lower panels of Fig.~\ref{fig:appendix}, the $\psi$-basis occupation number evolves smoothly across $p/H_{\rm inf}=\xi$ and remains well behaved in the small-mass regime. The sharp jump and subsequent saturation observed in the $Y$ basis should therefore not be interpreted as physical fermion production. Rather, they are artifacts of the instantaneous particle-number definition in the $Y$ basis, where the positive- and negative-frequency eigenstates are effectively exchanged when $\tilde k_-$ crosses zero. This interpretation is also consistent with the fact that the axion--fermion production effect should vanish in the massless limit. We therefore use the $\psi$-basis occupation number as the physically well-behaved definition in the small-mass regime.

The behavior of the occupation number can be understood directly from the structure of the Bogolyubov coefficient. In the limit $\mu\to0$, the Bogolyubov coefficient $\tilde{\beta}_r$ can be written as:
\begin{equation}\label{Y_BC2}
\begin{aligned}
\tilde{\beta}_r(t)
=
\frac{e^{ir\varphi_{\mathbf k}/2}}{2}
\left[
\sqrt{1+\text{Sign}\left[\tilde{k}_r(t)\right]}\,\tilde u_r(t)
-
r\sqrt{1-\text{Sign}\left[\tilde{k}_r(t)\right]}\,\tilde v_r(t)
\right],
\end{aligned}
\end{equation}
where Sign denotes the sign function. For the positive-helicity mode, $\tilde k_+$ is always positive. Therefore, $|\tilde\beta_+|^2$ is always dominated by $|\tilde u_+|^2\propto \mathcal{O}(\mu^2)$, and no unphysical behavior appears. For negative-helicity modes, however, one has $\tilde k_-<0$ at early times, so that $|\tilde\beta_-|^2$ is dominated by $|\tilde v_-|^2\propto \mathcal O(\mu^2)$. As the physical momentum is redshifted, $\tilde k_-$ crosses zero. In the strict $\mu\to0$ limit, the expression for $|\tilde \beta_-|^2$ becomes singular at $\tilde k_-=0$. After the crossing, when $\tilde k_->0$, $|\tilde \beta_-|^2$ becomes dominated by $|\tilde u_-|^2\propto \mathcal O(1)$, leading to an apparent jump from an $\mathcal O(\mu^2)$ occupation number to an $\mathcal O(1)$ value. This would imply nearly complete particle occupation even in the limit where the axion-fermion coupling vanishes, demonstrating that the result is manifestly unphysical and originates from the breakdown of the $Y$ basis particle definition rather than from genuine particle production.

The essential reason for the unphysical result is that the instantaneous occupation number defined in the $Y$ basis is not a physically appropriate particle-number definition in this regime. In the small-mass regime, its value is highly sensitive to the sign of $\tilde k_r$, since the instantaneous positive- and negative-frequency eigenstates are exchanged when $\tilde k_r$ crosses zero. By contrast, the fermion occupation number defined in the $\psi$ basis remains well behaved at all times and does not suffer from this singular behavior, as can be seen from Eq.~\eqref{eq:alpha_beta_inflation}. Therefore, we perform the calculation in the $\psi$ basis.


\bibliographystyle{JHEP}
\bibliography{biblio.bib}

\end{document}